\documentclass[conference]{IEEEtran}
\IEEEoverridecommandlockouts
\usepackage{cite}
\usepackage{multicol}
\usepackage{subcaption}
\usepackage{etoolbox}
\usepackage{relsize}
\usepackage{empheq}

\usepackage{amsmath,amssymb,amsfonts}
\usepackage[ruled,vlined]{algorithm2e}
\usepackage{algorithmic}
\usepackage{graphicx}
\usepackage{textcomp}
\usepackage{xcolor}
\usepackage{gensymb}
\usepackage{siunitx}
\def\BibTeX{{\rm B\kern-.05em{\sc i\kern-.025em b}\kern-.08em
    T\kern-.1667em\lower.7ex\hbox{E}\kern-.125emX}}
    
\usepackage[utf8]{inputenc}
\usepackage{amsmath}
\usepackage{braket}
\usepackage{graphicx}
\usepackage{amsfonts}
\usepackage[section]{placeins} 
\usepackage{pgfplots}
\usepackage{csquotes}
\usepackage{hhline}
\usepackage{amssymb}
\usepackage{listings}
\usepackage{color}
\definecolor{codegreen}{rgb}{0,0.6,0}
\definecolor{codegray}{rgb}{0.5,0.5,0.5}
\definecolor{codepurple}{rgb}{0.58,0,0.82}
\definecolor{backcolour}{rgb}{0.95,0.95,0.92}
\usepackage{tabularx}

\usepackage[utf8]{inputenc}
\usepackage{graphicx}
\usepackage{amssymb}
\usepackage{physics}
\usepackage{tikz}
\usepackage{bbold}
\usepackage{tikz}
\usetikzlibrary{quantikz}

\lstdefinestyle{mystyle}{
    backgroundcolor=\color{backcolour},   
    commentstyle=\color{codegreen},
    keywordstyle=\color{magenta},
    numberstyle=\tiny\color{codegray},
    stringstyle=\color{codepurple},
    basicstyle=\footnotesize,
    breakatwhitespace=false,         
    breaklines=true,                 
    captionpos=b,                    
    keepspaces=true,                 
    numbers=left,                    
    numbersep=5pt,                  
    showspaces=false,                
    showstringspaces=false,
    showtabs=false,                  
    tabsize=2
}

\usepackage{subcaption}

\usepackage{dcolumn}
\usepackage{tabularx}
\usepackage[colorlinks=true,linkcolor=blue,citecolor=blue,urlcolor=blue]{hyperref}
\usepackage{float}
\newcolumntype{C}{>{\centering\arraybackslash}X}
\pgfplotsset{compat=1.15}
\usepackage{multirow}
\begin{document}

\title{Investigation of Quantum Support Vector Machine for Classification in NISQ era}

\author{

\IEEEauthorblockN{1\textsuperscript{st} Anekait Kariya}
\IEEEauthorblockA{\textit{Department of Computer Science, BITS Pilani Goa Campus,} \\ \textit{Zuarinagar, Sancoale, 403726, Goa, India}\\ \textit{Bikash's Quantum (OPC) Pvt. Ltd.}\\ \textit{Balindi, Mohanpur, 741246,}\\ \textit{Nadia, West Bengal, India}\\ \textit{f20170031@goa.bits-pilani.ac.in}}

\and

\IEEEauthorblockN{2\textsuperscript{nd} Bikash K. Behera}
\IEEEauthorblockA{\textit{Bikash's Quantum (OPC) Pvt. Ltd.}\\ \textit{Balindi, Mohanpur, 741246,}\\\textit{Nadia, West Bengal, India}\\
\textit{bikas.riki@gmail.com}}
}

\maketitle

\begin{abstract}
Quantum machine learning is at the crossroads of two of the most exciting current areas of research; quantum computing and classical machine learning. It explores the interaction between quantum computing and machine learning, investigating how results and techniques from one field can be used to solve the problems of the other. Here, we investigate quantum support vector machine (QSVM) algorithm and its circuit version on present quantum computers. We propose a general encoding procedure extending QSVM algorithm, which would allow one to feed vectors with higher dimension in the training-data oracle of QSVM. We compute the efficiency of the QSVM circuit implementation method by encoding training and testing data sample in quantum circuits and running them on quantum simulator and real chip for two datasets; 6/9 and banknote. We highlight the technical difficulties one would face while applying the QSVM algorithm on current NISQ era devices. Then we propose a new method to classify these datasets with enhanced efficiencies for the above datasets both on simulator and real chips.
\end{abstract}

\maketitle
\section{Introduction}

In 1981, Feynman proposed the idea of using quantum computers to simulate quantum systems that are too hard for classical computers \cite{FeynmanIJTP1982}. 1992 on-wards quantum algorithms such as Deutsch-Jozsa algorithm \cite{DeutschPRSA1985}, Shor's algorithm \cite{ShorSIAM1999}, Grover's algorithm \cite{GroverPRL1997}, and many more have been proposed. After claims of significant theoretical speedup by these algorithms compare to classical methods, in 1998 first ever demonstration of quantum algorithm on a working quantum computer was shown \cite{ChuangPRL1998, JonesJCP1998}. First of it's kind quantum computers were based on the phenomenon of nuclear magnetic resonance (NMR) \cite{LloydScience1993,DiVincenzoScience1995,CoryPNAS1997,GershenfeldScience1997}. Soon, different architectures such as trapped ion \cite{Pogorelov,CiracPRL1995}, spin \cite{LossPRA1998}, silicon chip based \cite{VeldhorstNC2017}, diamond based \cite{GreentreeMT2008, PezzagnaAPR2021}, superconducting-qubit based \cite{DiCarloNature2009,BarendsNature2014,HongPRA2020}, photonics-qubit based \cite{KnillNature2001} have been developed for a quantum computer. In the recent years, massive progress has been made both experimentally and theoretically in demonstrating the potential of quantum computing \cite{PreskillQ2018,BoixoNaPh2018}. One such feat was marked by Google in 2019, by achieving quantum supremacy using a programmable 53-qubit quantum computer \cite{AruteNAT2019}. This caught tremendous attention from the scientific community all over the world, encouraging researchers from various disciplines like computer science, electrical engineering, chemistry, biology and in recent times even finance to look for use cases in quantum computers. This has also been facilitated by number of platforms such as IBM quantum experience (IBM QE) \cite{IBMQE}, D wave \cite{D-Wave}, Amazon braket \cite{Amazon}, intel \cite{Intel}, Microsoft \cite{Azure}, Rigetti \cite{Rigetti}, Xanadu \cite{Xanadu}, Google AI \cite{GoogleAI} and a few others that allow online access to their quantum computers. Among these, IBM QE is a free online based platform where users can run quantum circuits on up to 5000-qubits simulator and 5-qubit real quantum chips \cite{IBMQE}. Availability of such platforms, was an exciting step for young researchers, who could now have fun on expensive quantum computers via cloud; in pursuit of finding problems for which quantum computation would be advantageous over classical means. Since the reduced entry barrier in the field, millions of quantum computational and quantum informational tasks have been run on such platforms. These tasks included but not limited to; quantum algorithm \cite{LeeSciRep2019,SrinivasanarXiv2018, LiQMQM2017}, quantum communication \cite{BeheraQIP2019,BKBQIP2019}, quantum cryptography \cite{MubayedhICCAIS2019}, quantum machine learning \cite{ParkarXiv2020,YangarXiv2019}, quantum simulations \cite{MahantiQIP2019}, quantum robotics \cite{MishraQIP2020}, and quantum games \cite{PalEPL2019} to name a few.

Machine learning and artificial intelligence has been transforming our lives since several decades. Today technological companies like Amazon, Google, Netflix, Facebook, Apple, and many more use machine learning as the core idea. These companies benefit heavily from the range of applications machine learning has to offer, like product recommendation systems, image and speech recognition, spam detection systems, self-driving cars, etc. Machine learning thrives to learn patterns in the data in-order to achieve these tasks \cite{SchriderTiG2018}. A machine learning task is generally of two categories supervised learning \cite{KotsiantisAIR2006} and unsupervised learning \cite{GhahramaniLNAI2004}. Supervised learning refers to learning patterns from a labelled dataset, to be able then predict the label of a new data-point. Few examples of supervised learning are linear and logistic regression, support vector machine \cite{CortesML1995}, k-nearest neighbors. Where as in unsupervised learning the dataset isn't labelled and the task is to find patterns and structure in the data. Few examples of unsupervised learning include identifying clusters \cite{RuspiniInCo1969}, principal component analysis (PCA) \cite{WoldCILS1987}. Machine learning algorithms rely heavily on data to able to make predictions for a new data-points, but today the size of datasets have been increasing rapidly. It is estimated that global data is in the order of zettabytes $10^{21}$ \cite{statsGlobaldata}, which has started to become a limit to be processed by classical computers. In the light of the increasing global data, researchers from all over the world are turning to quantum computing to provide a solution in the near future. This has given rise to the field of quantum machine learning and quantum artificial intelligence. The hope to be able to process large amounts of data arises from the concept of quantum random access memory (QRAM) \cite{GiovannettiPRL2008}. In quantum machine learning, encoding classical data in quantum mechanical form is an important problem and many encoding procedure are proposed. In one such encoding procedure, of amplitude encoding it requires only $\mathcal{O}(\log d)$ qubits as compare to $ \mathcal{O}(d) $ bits for classical case, leading to a exponential compression in representation of data \cite{LloydYT2016}. This is basis for speedup in quantum version for the following methods.

\begin{table}[H]
    \caption{\textbf{Computational complexity comparison; \\ classical versus quantum} \cite{LloydYT2016,BiamonteNat2017}}
    \begin{tabular}{|>{\centering}p{3.5cm}|>{\centering}p{2.2cm}|>{\centering}p{1.8cm}|}
        \hline
        Methods; Algorithm & Classical & Quantum \tabularnewline
        \hline
         Fast fourier transform (FFT) & $\mathcal{O}(d\log{d})$ & $\mathcal{O}((\log{d})^2)$  \tabularnewline
         \hline
         Eigenvectors and eigenvalues & $\mathcal{O}(d^3)$ $\mathcal{O}(sd^2)$ & $\mathcal{O}((\log{d})^2)$ \tabularnewline
         \hline
         Matrix inversion $\vec{x}=A^{-1}\vec{b}  $ & $\mathcal{O}(d\log{d})$ & $\mathcal{O}((\log{d})^3)$ \tabularnewline 
         \hline
    \end{tabular}
    \label{QML_table}
\end{table}

In this paper we focus upon the quantum support vector machine (QSVM) algorithm given by Rebentrost \emph{et al.} \cite{RebentrostPRL2014} in 2014. At the heart of this algorithm lies the non-sparse matrix exponentiation technique, i.e. the HHL algorithm given by Harrow \textit{et al.} \cite{HarrowPRL2009} in 2009. We explore the depths of the QSVM algorithm and provide an implementation for it on qiskit. To help devise the quantum circuit version of QSVM algorithm we refer to the experimental demonstration of QSVM done by Li \emph{et al.} \cite{LiPRL2015} on a four-qubit nuclear magnetic resonance (NMR) in 2015. This experimental demonstration was shown for a 6/9 hand-written digit recognition task, where each datapoint had two features. The task is an optical character recognition (OCR) problem of binary classification of digits 6 and 9, in which standard font (times new roman) and handwritten characters are taken as training and testing samples respectively. We implement this task on today's current state of art quantum computer using the IBM quantum experience \cite{IBMQE} platform; for both simulator and real chip. Then, we take the bank note authentication dataset having four features, for which we propose an encoding procedure in a two-qubit system, extending the QSVM implementation. Our proposed encoding technique could also be used to feed a general four features vector in the QSVM circuit, compared to just two features in the 6/9 case. We apply two classical pre-processing methods in the case of bank note authentication data to find the training vectors. One is simply using averages of features of the training data and other one is to use k-means to cluster the data in two classes and then use the center of the clusters as the training vector. Using the IBM quantum experience \cite{IBMQE} platform, we run the QSVM algorithm for banknote dataset on both simulator and real chip; for both averages and k-means. In the last section of this paper inspired by following works we propose a different method for calculating the inner product as compared the QSVM algorithm. We notice that this method has much shorter circuit and performs with better accuracy on both the datasets. To conclude, we highlight the limitation of the QSVM algorithm on today's NISQ era quantum computers and refer to other caveats \cite{AaronsonNatP2015}, which we must solve before we can completely harness the exponential speedup claimed to be offered by quantum algorithms.

The rest of the paper is organized as follows. In Sec. \ref{Sec2}, we discuss the quantum support vector machine algorithm \cite{RebentrostPRL2014} and it's general circuit. Sec. \ref{Sec3} applies the QSVM algorithm to 6/9 digit dataset \cite{LiPRL2015}. In Sec. \ref{Sec4}, we apply the QSVM algorithm to banknote dataset, discuss the technical difficulties we faced and propose an encoding procedure as well as two classical pre-processing methods to overcome them. In Sec. \ref{Sec5}, we propose a new method and apply it on both 6/9 and banknote dataset. We observe this method performs better compare to QSVM circuit implementation. Finally, we conclude in discussion and future directions of work.

\section{Quantum Support Vector Machine \label{Sec2}}
 \begin{figure*}[!ht]
\begin{subfigure}{0.54\linewidth}
\includegraphics[width=\linewidth]{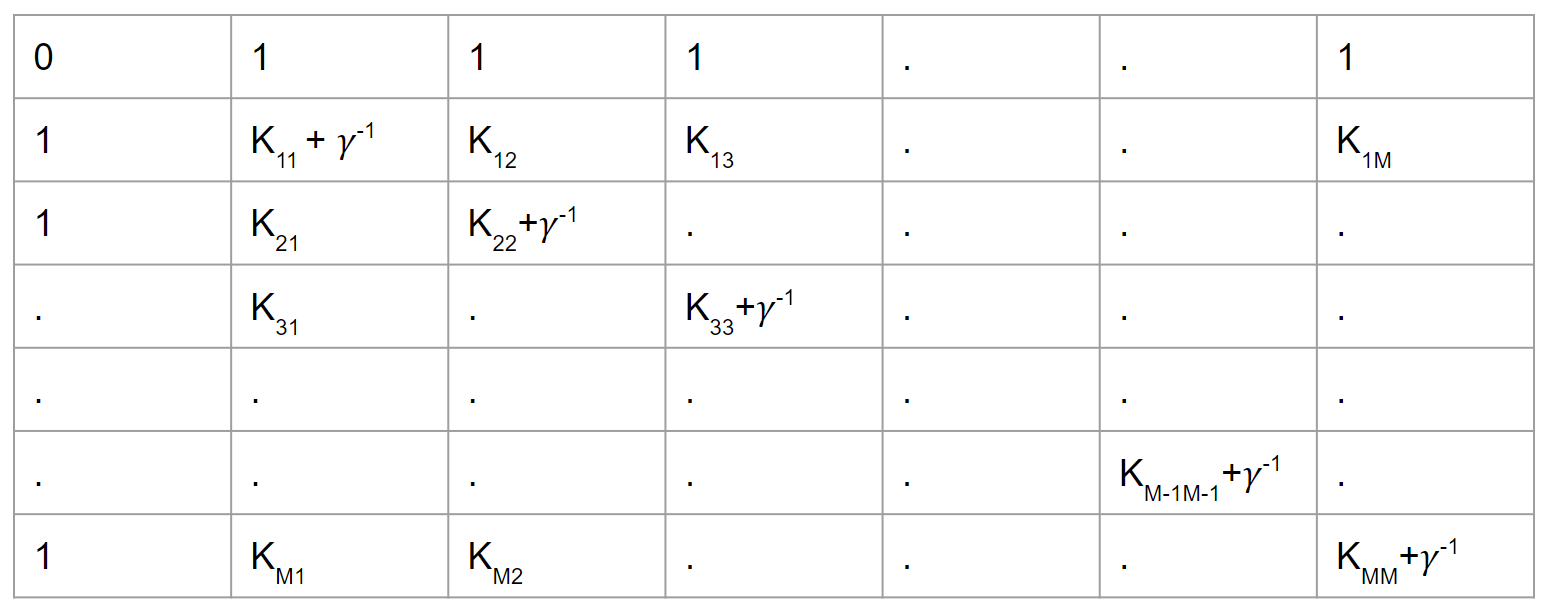}
\caption{\textbf{General F matrix}. }
\label{QSVM_Fig1a}
\end{subfigure}
\begin{subfigure}{0.45\linewidth}
\includegraphics[width=\linewidth]{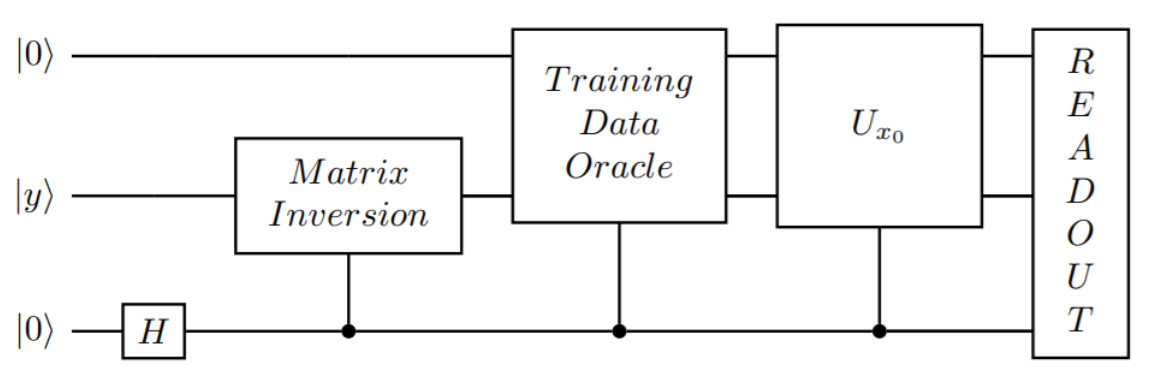}
\caption{\textbf{QSVM algorithm overview with readout \cite{LiPRL2015}}. }
\label{QSVM_Fig1b}
\end{subfigure}
\begin{subfigure}{0.36\linewidth}
\includegraphics[width=\linewidth]{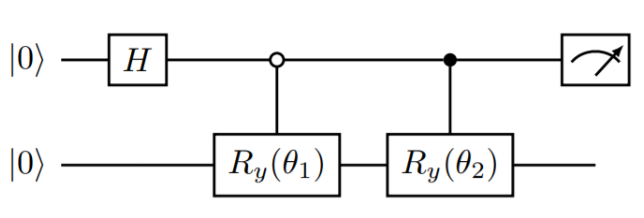}
\caption{\textbf{QSVM training-data oracle \cite{LiPRL2015}}. }  
\label{QSVM_Fig1c}
\end{subfigure}
\begin{subfigure}{0.63\linewidth}
\includegraphics[width=\linewidth]{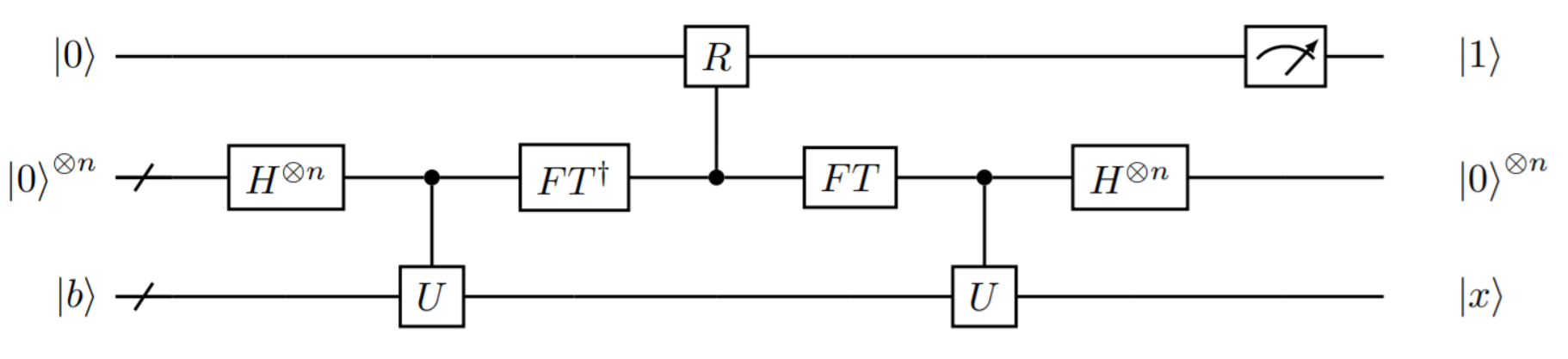}
\caption{\textbf{QSVM matrix inversion - HHL algorithm \cite{DuttaIET2020}.} }
\label{QSVM_Fig1d}
\end{subfigure}
\begin{subfigure}{0.99\linewidth}
\includegraphics[width=\linewidth]{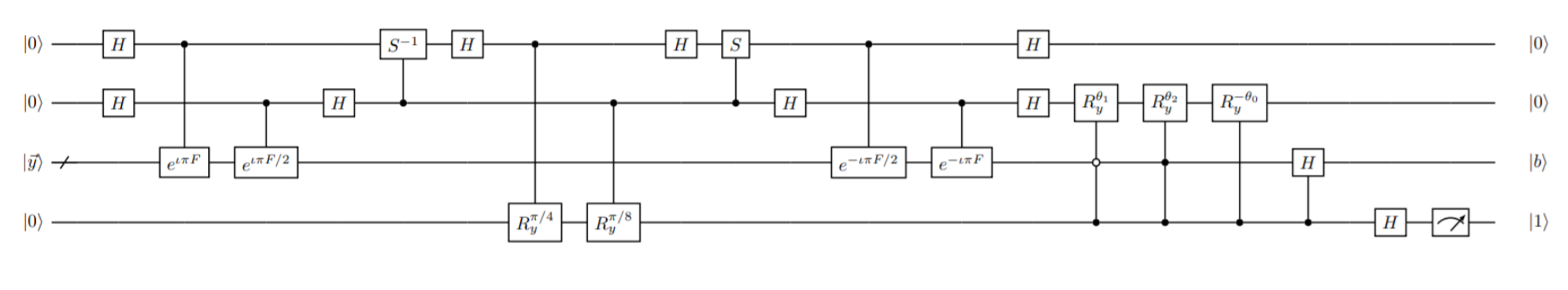}
\caption{\textbf{QSVM general circuit \cite{LiPRL2015}.}}
\label{QSVM_Fig1e}
\end{subfigure}\hfill
\caption{ \textbf{QSVM algorithm: } \textbf{(a)} F is a $(M+1) \times (M+1)$ matrix with the essential part being the kernel matrix K, where $K_{i,j} = \kappa(\Vec{x_i}, \Vec{x_j})$. The user-specified weight $\gamma$ determines the relative weight of the training error and the SVM objective \cite{SuykensNPL1999}. {\textbf{(b)} Here, we use three qubits; the first one stands for the training register, the second one is used to input $|y\rangle$ state, and the third one is the ancilla qubit. After the operation of matrix inversion, training data oracle, and unitary operation, measurement is performed. \textbf{(c)} The quantum circuit creates the training data oracle state, then after performing measurement on the ancilla qubit, the inner product between the training and the testing data is calculated. \textbf{(d)} F matrix inversion is calculated after applying the phase estimation algorithm, rotation operator, and un-compute process as given in the circuit. \textbf{(e)} This is the general circuit comprising all the above methods, starting from the initial state $\ket{0,\Vec{y}}$, which is converted to the state $\ket{b,\Vec{\alpha}}$ on matrix inversion of F, which then introduced to training-data oracle converts to $\ket{\Vec{u}}$. To query our algorithm lastly we perform the inverse $U_{x0}$ operation, after then measurement of ancilla qubit is used for classification of the query state $\Vec{x_0}$.}}
\label{QSVM_Fig1}
\end{figure*}

Support Vector Machine (SVM) is a supervised classical machine learning model \cite{KotsiantisAIR2006}, used for binary classification of new testing vectors. SVM is a class of problems where the number of equations are more than the number of unknowns, which is called a over-determined system of equations. Though there are many approaches proposed by the researchers, the method of least-squares fitting \cite{SuykensNPL1999} is a standard approach to approximately solve such a regression analysis problem, that minimizes the sum of the squares of the residuals made in every single equation. SVM is a supervised algorithm which aims to learn from the training samples in-order to classify a new data sample into positive or negative class. It constructs hyper-plane with $\Vec{w}.\Vec{x} + b = 0$ such that $\Vec{w}.\Vec{x} + b \geq 1$ for a training point $\Vec{x}_i$ in the positive class, and $\Vec{w}.\Vec{x} + b \leq -1 $ for a training point $\Vec{x}_i$ in the negative class. During the training process, the algorithm aims to maximizes the gap between the two classes, which is intuitive as we want to separate two classes as far as possible, in-order to get a sharper estimate for classification result of new data sample like $\Vec{x_0}$. Mathematically we can see objective of SVM is to find hyper-plane that maximize the distance $2/ |\Vec{w}|$ constraint to  $\Vec{y_i}(\Vec{w}.\Vec{x_i} + b) \geq 1$. The normal vector $\Vec{w}$ can be written as $\Vec{w} = \sum_{i=1}^{M} \alpha_i \Vec{x}_i$ where $\alpha_i$ is the weight of the $i^{th}$ training vector $\Vec{x}_i$. Thus, obtaining optimal parameters b and $\alpha_i$ is same as finding the optimal hyper-plane. To classify the new vector, is analogous to knowing which side of the hyper-plane it lies, i.e., $ y_i(\Vec{x}_0) = sgn(\Vec{w}.\Vec{x} + b) $. After having the optimal parameters, classification now becomes a linear operation. From the least-squares approximation of SVM the optimal parameters can be obtained by solving a linear equation,

\begin{eqnarray}
\Vec{F}(b,\alpha_1,\alpha_2,\alpha_3,...,\alpha_M)^T = (0,y_1,y_2,y_3,...y_M)^T.
\label{QSVM_Eq1}
\end{eqnarray}

General form of F can be observed in Fig. (\ref{QSVM_Fig1a}), where we adopt the linear kernels $K_{i,j} = \kappa(\Vec{x}_i,\Vec{x}_j) = \Vec{x}_i. \Vec{x}_j $. Thus to find the hyper-plane parameters we use matrix inversion of F : $(b,{\vec{\alpha}_i}^T)^T  = {\tilde F^{ - 1}} (0,{\vec{y}_i}^T)^T$.

Quantum counterpart of the training data $\Vec{x}_i$ is the state $ \ket{\Vec{x}_i} = 1/|\Vec{x}_i| \sum_{j=1}^{N}(\Vec{x}_i)_j \ket{j}$. For a case of 2 training points, represented by parameter $\theta_1$ and $\theta_2$ the training-data oracle would be as in Fig. (\ref{QSVM_Fig1c}). In general for M training points, training-data could be used to prepare the state $ \ket{\chi} = 1/\sqrt {{N_\chi }} \sum_{i = 1}^M |\Vec{x}_i| \ket{i} \ket{\vec x} $, with $ N_{\chi} = \sum_{i = 1}^M |\Vec{x}_i|^2$ from the initial state $1/\sqrt M \sum_{i = 1}^M \ket{i}$. To obtain kernel matrix $\textit{K}$, in quantum counterpart, the quantum density matrix is reduced to the kernel matrix $\textit{K}$, after discarding the training set registers, up to a constant factor $\textbf{tr}\textit{K}$ \cite{LiPRL2015}. The quantum registers are first initialized into $\ket{0,\Vec{y}}$ which after matrix inversion of F is transferred to the quantum state $\ket{b,\Vec{\alpha}}$.

\begin{eqnarray}
\ket{0,\Vec{y}} &=& 1/\sqrt {{N_{0,y} }} (\ket{0} + \sum_{i = 1}^M \Vec{y}_i \ket{i} \nonumber \\
\ket{b,\Vec{\alpha}} &=& 1/\sqrt {{N_{b,\alpha} }} (b\ket{0} + \sum_{i = 1}^M \Vec{\alpha}_i \ket{i} )
\label{QSVM_Eq2}
\end{eqnarray}

Here, $N_{b,\alpha} $ and $N_{0,y}$ are normalization factors. With the optimized parameters b and $\alpha_i$, the classification results $y(\Vec{x}_0)$ can be represented by,

\begin{eqnarray}
y(\Vec{x}_0) = sgn( \sum_{i = 1}^M \alpha_i( \Vec{x}_i.\Vec{x}_0 ) + b ) 
\label{QSVM_Eq3}
\end{eqnarray}

\begin{figure*}
\begin{subfigure}{0.27\linewidth}
\includegraphics[width=\linewidth]{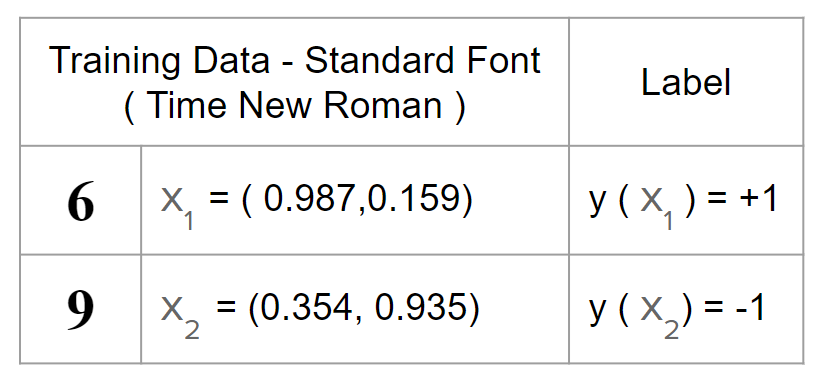}
\caption{\textbf{Training dataset \cite{LiPRL2015}}. }
\label{QSVM_Fig2a}
\end{subfigure}
\begin{subfigure}{0.31\linewidth}
\includegraphics[width=\linewidth]{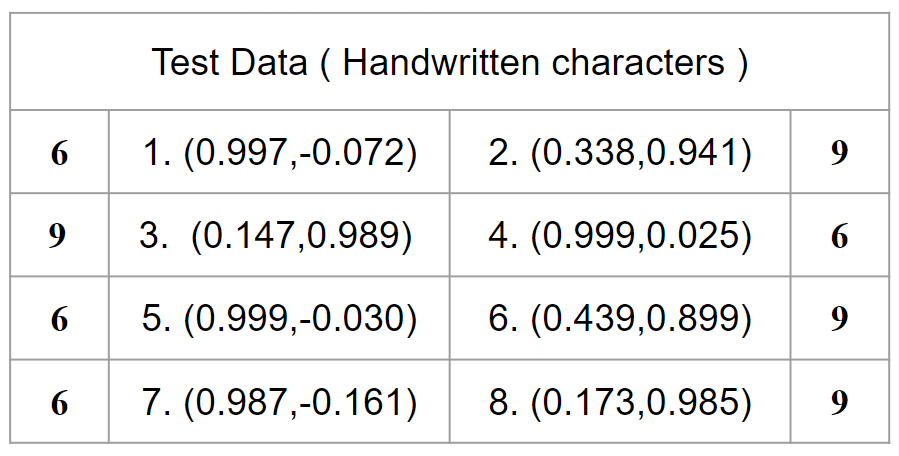} 
\caption{\textbf{Testing dataset \cite{LiPRL2015}}. }
\label{QSVM_Fig2b}
\end{subfigure}
\begin{subfigure}{0.42\linewidth}
\includegraphics[width=\linewidth]{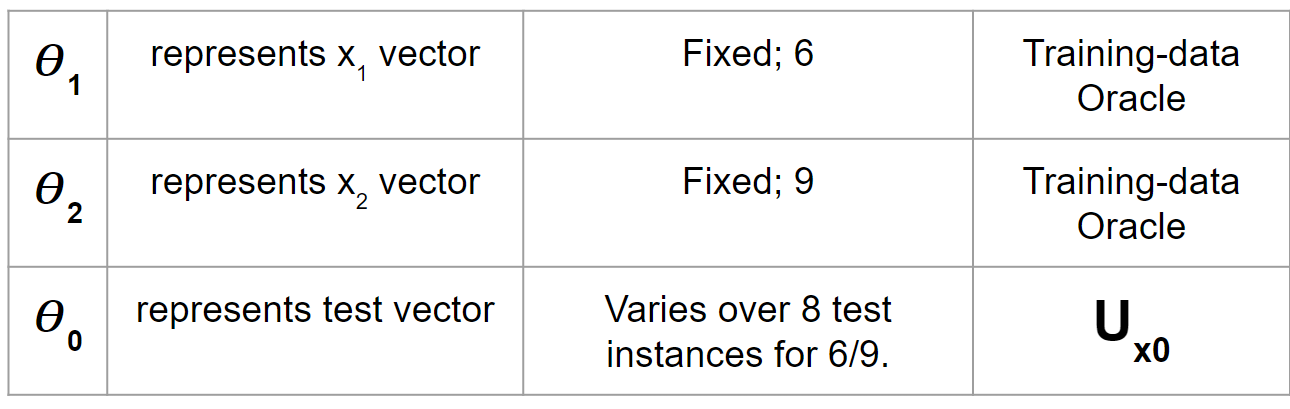}
\caption{\textbf{Theta description table}. }
\label{QSVM_Fig2c}
\end{subfigure}
\begin{subfigure}{0.36\linewidth}
\includegraphics[width=\linewidth]{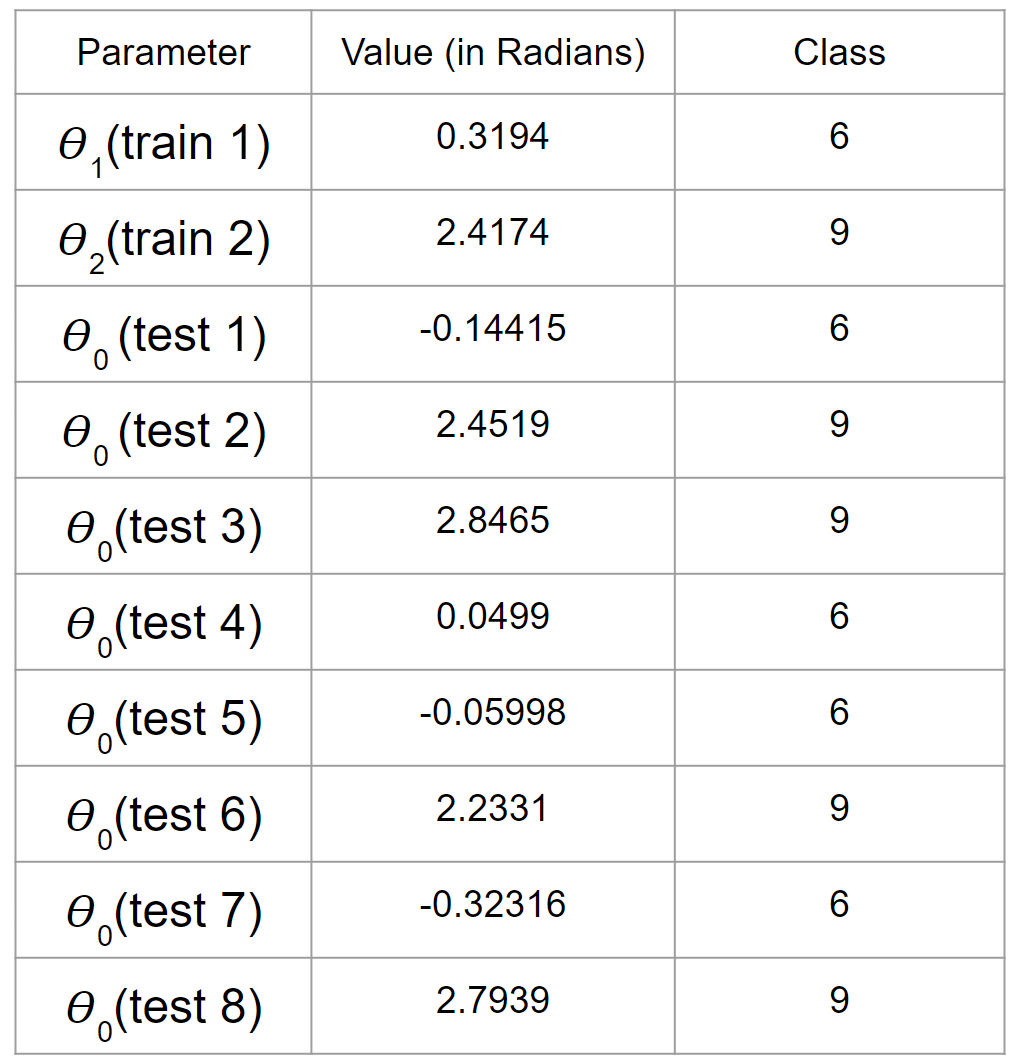}
\caption{\textbf{QSVM theta values table}. }
\label{QSVM_Fig2d}
\end{subfigure}
\begin{subfigure}{0.24\linewidth}
\includegraphics[width=\linewidth]{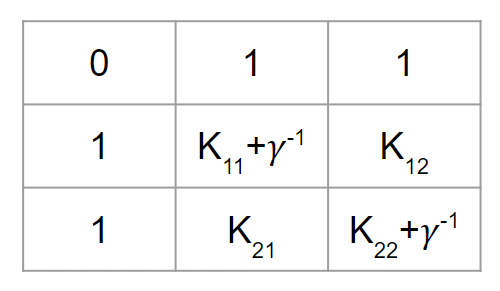} 
\caption{\textbf{General F: M=2 }. }
\label{QSVM_Fig2e}
\end{subfigure}
\begin{subfigure}{0.18\linewidth}
\includegraphics[width=\linewidth]{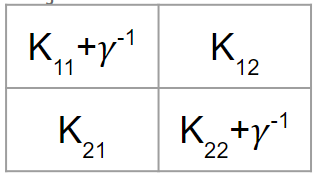} 
\caption{\textbf{F: M=2; b=0}. }
\label{QSVM_Fig2f}
\end{subfigure}
\begin{subfigure}{0.194\linewidth}
\includegraphics[width=\linewidth]{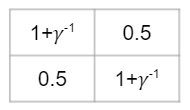} 
\caption{\textbf{F: 6/9 case}.}
\label{QSVM_Fig2g}
\end{subfigure}
\caption{\textbf{6/9 dataset overview:} \textbf{(a)} Values of the training vectors corresponding to characters 6 and 9, in their standard font, i.e. times new roman are shown. \textbf{(b)} Arbitrarily chosen handwritten samples are used as testing sample. Testing dataset table describes the vector values of eight test points labelled as test 1 to test 8. After training of the algorithm, these eight test points are used to check the algorithm's accuracy for 6/9 case. \textbf{(c)} Here, we define $\theta_{1}$, $\theta_{2}$ and $\theta_{0}$ which correspond to the parameters $R{y}^{\theta_{1}}$, $R{y}^{\theta_{2}}$ and $R{y}^{-\theta_{0}}$ respectively, of the QSVM general circuit (Fig. \ref{QSVM_Fig1e}). $R{y}^{\theta_{1}}$, $R{y}^{\theta_{2}}$ form part of the QSVM training-data oracle (Fig. \ref{QSVM_Fig1c}) and $R{y}^{-\theta_{0}}$ form part of the unitary operator $U_{x_0}$ (Fig. \ref{QSVM_Fig1b}). \textbf{(d)} This table links the parameters $R{y}^{\theta_{1}}$, $R{y}^{\theta_{2}}$, $R{y}^{-\theta_{0}}$ to their corresponding values. These values are calculated using the expression (Eq. \eqref{QSVM_Eq5}) and using the features from the dataset tables from Fig. \ref{QSVM_Fig2a} and Fig. \ref{QSVM_Fig2b}. \textbf{(e)} F matrix for M = 2 follows directly from the general F matrix (\ref{QSVM_Fig1a}), it is a $3\times 3$ matrix. \textbf{(f)} F matrix in this case is a $2\times 2$ matrix, which follows from setting non-offset reduction to zero, i.e., b = 0. \textbf{(g)} F matrix for 6/9 dataset follows from substituting the $K_{11}$, $K_{12}$, $K_{21}$, $K_{22}$ values for the 6/9 case.}
\label{QSVM_Fig2}
\end{figure*}

This could be reproduced by the overlap of the two quantum states: $y(\Vec{x}_0) = sgn( \bra{\Vec{v}_0}\ket{\Vec{u}})$ with the training-data state and the query state,

\begin{eqnarray}
 \ket{\Vec{u}} &=& 1/\sqrt {N_{\Vec{u} }} [b\ket{0}\ket{0} + \sum_{i = 1}^M abs(\Vec{x}_i)\Vec{\alpha}_i\ket{i}\ket{\Vec{x}_i} ] \nonumber \\  
 \ket{\Vec{v}_0} &=& 1/\sqrt {N_{\Vec{x}_0 }} [\ket{0}\ket{0} + \sum_{i = 1}^M abs(\Vec{x}_0)\ket{i}\ket{\Vec{x}_0} ]
 \label{QSVM_Eq4}
\end{eqnarray}

Here, the training-data state $\ket{\Vec{u}}$ could be easily obtained by calling the training-data oracle (Fig. \ref{QSVM_Fig1b}) on $\ket{b,\Vec{\alpha}} $ and an additional training register (Fig. \ref{QSVM_Fig1a}). To introduce the information of the query vector $\Vec{x}_0$, a unitary inverse operation $U_{x_{0}}$ is applied to transfer $\ket{\Vec{v}_0}$ to $\ket{00}$, i.e., $U_{x_{0}} \ket{\Vec{v}_0} = \ket{00}$. Then, the expansion coefficients $\bra{00} U_{x_{0}} \ket{\Vec{u}} = \bra{v_0}\ket{u}$ produce the classification result $y(\Vec{x}_0)$. It is noted that the unitary operations are conditional operations controlled by an ancillary qubit. Hence, the final state will be $\ket{\Psi} = \ket{\Phi} \ket{1}_A + \ket{00}\ket{0}_A$ denoting the states of the ancillary qubit. By measuring the expectation value of the coherent term $O = \bra{00} \ket{00} * (\bra{1}\ket{0})_A$ the classification result will be revealed by $y(\vec{x}_0)$. The matrix inversion problem is also called the HHL algorithm, which was proposed by Harrow \emph{et al.} \cite{HarrowPRL2009} and below is a systematic view of it's circuit implementation \cite{DuttaIET2020}, which will be useful for us. Here, we use the HHL algorithm for solving linear equations. The quantum circuit given in Fig. (\ref{QSVM_Fig1c}) is the HHL algorithm, whose detail mathematics and general form can be found in Ref. \cite{DuttaIET2020}. It maps the state $\ket{0,\Vec{y}}$ to $\ket{b,\Vec{\alpha}}$. In Fig. (\ref{QSVM_Fig1}), we highlight the details of the QSVM algorithm which we require for the scope of this paper. For further details of the QSVM algorithm, like the steps involved in the HHL algorithm, kernel matrix approximation, error analysis, optimality and more; enthusiastic readers could refer to these papers \cite{HarrowPRL2009,RebentrostPRL2014,LiPRL2015,DuttaIET2020}.

\section{6/9 dataset \label{Sec3}}
\subsection{Overview of 6/9 dataset}\label{Sec3A}
\begin{figure*}
\includegraphics[width=\textwidth]{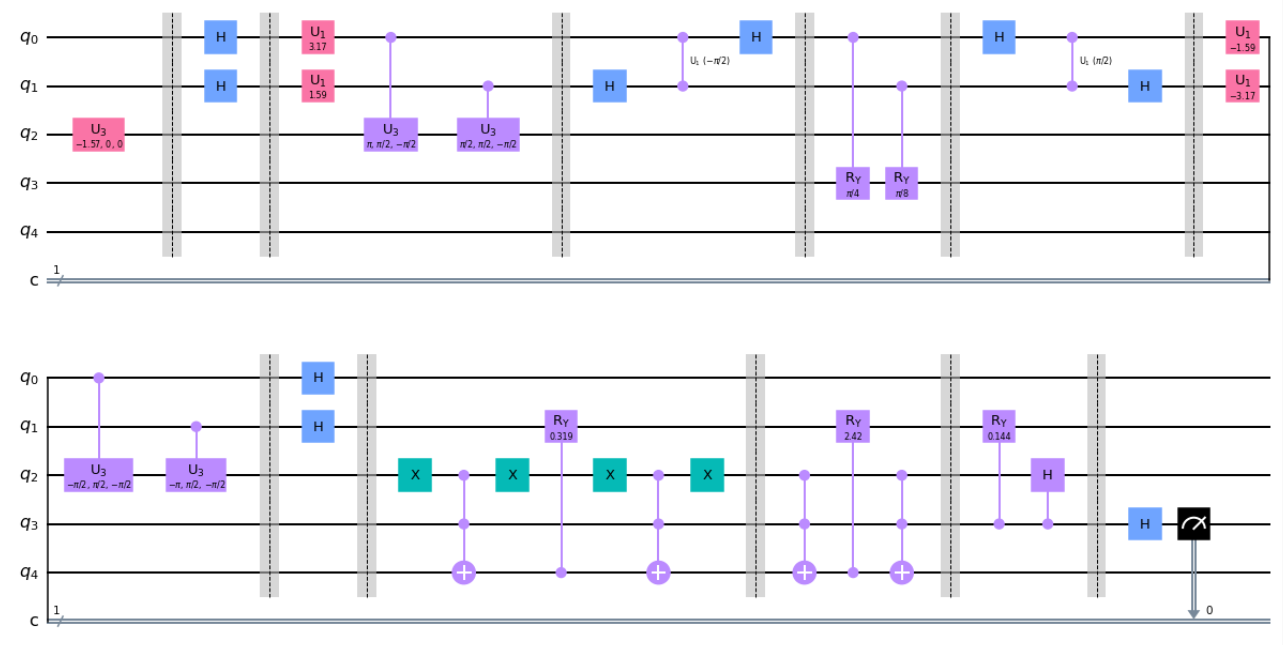}
\caption{\textbf{QSVM circuit for 6/9 : } The above circuit was run on IBM qasm simulator and IBM belem to obtain our 6/9 binary classification results. It was obtained by assigning all the parameters values to the general QSVM circuit (Fig. \ref{QSVM_Fig1e}). It was implemented 8 times for 8 different test points by changing only $R{y}^{-\theta_{0}}$ whose values we get from Fig. \ref{QSVM_Fig2d}. Here, we show the circuit only for test 1; $\theta_{0}$ = -0.144, i.e. $\theta$ corresponding to $R{y}^{-\theta_{0}}$ = +0.144.}
\label{QSVM_Fig3}
\end{figure*}

In this section, we would like to apply the QSVM algorithm to the popular optical character recognition (OCR) problem \cite{TrierPatRe1996}, which is to classify handwritten characters in a candidate set. There are many sub classes of the OCR problem, but to study the proof of concept for the QSVM algorithm, we restrict ourselves to binary classification of digits ``6" and digit ``9" \cite{LiPRL2015}. The two features for this 6/9 case, are horizontal and vertical ratios of the digits. To train the QSVM algorithm, we use the standard font (Times New Roman) of the characters 6 and 9. Training vectors can be seen in Fig. (\ref{QSVM_Fig2a}). To test the QSVM algorithm, we provide it with eight handwritten images of characters 6 or 9. Testing vectors can be seen in Fig. (\ref{QSVM_Fig2b}). Thus, our complete 6/9 dataset is given in Figs. (\ref{QSVM_Fig2a}) and (\ref{QSVM_Fig2b}).

\subsection{QSVM Circuit for 6/9}\label{Sec3B}

To apply QSVM general circuit (Fig. (\ref{QSVM_Fig1e})) to our 6/9 classification problem we need to find the parameters: F matrix, $R{y}^{\theta_{1}}$, $R{y}^{\theta_{2}}$ and $R{y}^{-\theta_{0}}$. Let us first find the parameters $R{y}^{\theta_{1}}$, $R{y}^{\theta_{2}}$ and $R{y}^{-\theta_{0}}$ as it follows directly from the 6/9 dataset. We define $\theta_{1}$, $\theta_{2}$ and $\theta_{0}$ specific to our 6/9 case, which is described in Fig. (\ref{QSVM_Fig2c}). $\theta_{1}$ and $\theta_{2}$ would correspond to training vectors (Fig. (\ref{QSVM_Fig2a})) and would be the part of the QSVM training-data oracle (Fig. (\ref{QSVM_Fig1c})). $\theta_{0}$ corresponds to the testing vectors (Fig. (\ref{QSVM_Fig2b})) and thus would be part of the unitary operator $U_{x_0}$ (Fig. (\ref{QSVM_Fig1b})). Next, to find the theta values ($\theta_{1}$, $\theta_{2}$, $\theta_{0}$) we can take the general form of vector $\vec{x} = (\alpha,\beta)$, where $\alpha$ and $\beta$ are the features of the vector and use the formulae.

\begin{eqnarray}
\theta = 2\tan^{-1}{\dfrac{\beta}{\alpha}}
\label{QSVM_Eq5}
\end{eqnarray}

which follows from the use of $R_y$ gates in the QSVM training-data oracle (Fig. (\ref{QSVM_Fig1c})) and $U_{x_0}$ (Fig. (\ref{QSVM_Fig1b})). Values for $\theta_{1}$ (train 1), $\theta_{2}$ (train 2) and $\theta_{0}$ (test 1 to 8) which ranges over eight test points; can be found in Fig. (\ref{QSVM_Fig1d}) as QSVM theta values table. This table will be used to feed the parameters $R{y}^{\theta_{1}}$, $R{y}^{\theta_{2}}$ and $R{y}^{-\theta_{0}}$ to the QSVM general circuit (Fig. (\ref{QSVM_Fig1e})) for 6/9 dataset. Now coming to F matrix, we have already seen the general F matrix in Fig. (\ref{QSVM_Fig1a}). This reduces to Fig. (\ref{QSVM_Fig2e}) as we have only 2 training vectors i.e., M=2. In our implementation, we set non-offset reduction, i.e., b = 0, converting the F matrix to Fig. (\ref{QSVM_Fig2f}). To finally get the F matrix for 6/9 case, we need to define and calculate the $K_{i,j}$ terms. Here, we use the dot product kernel, i.e., $K_{i,j} = \kappa(\Vec{x}_i,\Vec{x}_j) = \Vec{x}_i . \Vec{x}_j$. Using the features of test samples, from Fig. (\ref{QSVM_Fig2a}), we calculate $K_{1,2} = \kappa(\Vec{x}_1,\Vec{x}_2) = K_{2,1} = \Vec{x}_1 . \Vec{x}_2 = 0.5$. Similarly, $K_{1,1} = 1$ and $K_{2,2} = 1$. Thus F matrix for 6/9 case is found to be as in Fig. (\ref{QSVM_Fig2g}). Lastly, to feed this matrix to QSVM circuit we need to exponentiate the matrix. To do so, we first write the final F matrix in 6/9 case which we need to exponentiate as $F = {c_1}I + {c_2}X$. $e^{iF} = e^{i({c_1}I + {c_2}X)} = e^{i{c_1}I}e^{i{c_2}X} = e^{i{c_2}X}$, where $e^{i{c_1}I}$ is treated as a global phase. From Fig. (\ref{QSVM_Fig2g}), we can easily fill in the ${c_1}$ and ${c_2}$ values as ${c_1} = 1+ \gamma^{-1}$ and ${c_2} = 0.5$.

Thus, now we have all the parameters, which we can easily feed in the QSVM general circuit (Fig. (\ref{QSVM_Fig1e})). Finally the QSVM circuit for the 6/9 case is shown in Fig. (\ref{QSVM_Fig3}). To run this quantum circuit and all other quantum circuits in our paper we use the IBM quantum experience (IBM QE) \cite{IBMQE} platform. Currently, IBM QE \cite{IBMQE} allows access up to 5000 qubits on a simulator and up to 127 qubits on real chip. IBM offers simulator access to 5 types of system, namely; simulator statevector (32 qubits), qasm simulator (32 qubits), simulator extended stabilizer (63 qubits), simulator mps (100 qubits), simulator stabilizer (5000 qubits). IBM offers access to 23 real chip systems, which include 1, 5, 7, 16, 27, 65 and recently released 127 qubits system. Though the simulator access is freely available till 5000 qubits, on real chip it's only up to 5 qubits. Our QSVM circuit for the 6/9 case has 5 qubits, which allows us to run it on both simulator and real chip. 

\textbf{Results for 6/9 case:}
\begin{itemize}
    \item IBM qasm simulator; for 8192 shots, 6/9 circuit classifies 5/8 test samples, i.e., accuracy - \textbf{62.5\%}.
    \item IBM belem; for 8192 shots, 6/9 circuit classifies 5/8 test samples, i.e., accuracy - \textbf{62.5\%}.
\end{itemize}

We think these results could be explained by formation of erroneous solution state at the end of the HHL algorithm. Let the exact solution state expected at the end of the HHL algorithm be of the form of $\ket{\psi}=a\ket{0}+b\ket{1}$, but we believe instead a erroneous state, say of form $\ket{\psi_e}=c\ket{0}+d\ket{1}$ is being formed. Where as $a\neq c$, and $b\neq d$, thus the erroneous state doesn't have the expected probability amplitudes. So in the step following the HHL algorithm, we end up performing the control-anti-control and control-control rotational operators of the training data oracle (Fig. \ref{QSVM_Fig1c}) on a state which doesn't have the expected probabilities. Due to this, rotational operators are applied on this erroneous state, instead of being applied on $\ket{00}$ clock registers (which was expected). In the HHL algorithm, post selection is an important task, as without post-selection, the clock registers can not reset to $\ket{00}$ state, and the exact solution state is not achieved on the required qubit. Due to thus formed erroneous solution state, we do not expect accurate results even after applying controlled and anti-controlled rotational operators in the later step. 
Specification of quantum system used in this section.
\begin{enumerate}
    \item IBM qasm simulator: 32 qubits. Simulator type: General, context-aware.
    \item IBM belem specifications: 5 qubits; 16 quantum volume; 2.5K CLOPS; Avg. T1: 93.81 $\mu$s; Avg. T2: 102.93 $\mu$s.
\end{enumerate}

 

\section{Banknote dataset \label{Sec4}}
\subsection{Overview of banknote dataset \label{Sec4A}}
In this section, we extend the QSVM algorithm for banknote authentication dataset \cite{BNKaggle2018}, which is an important problem to identity if a given banknote is forged or real. Below we briefly mention the pre-processing involved to obtain the features for the banknote dataset \cite{BNKaggle2018} from the banknote specimen. Digitization of the dataset from images of banknote was done using industrial print inspection camera \cite{BNdatasetUCI2013}. The final images have $400\times400$ pixels and gray-scale pictures with a resolution of about 660 dpi. Wavelet transform \cite{XizhiMMIT2008} tool were used to extract features from the images. The complete dataset has 1372 rows and 5 columns. Fig. \ref{QSVM_Fig4a} shows the head of the dataset, i.e., the first five rows. The dataset has four features, namely variance of wavelet transformed image, skewness of wavelet transformed image, kurtosis of wavelet transformed image, entropy of image, all four of which are continuous. Table \ref{QSVM_table} describes these four features in more detail. Class is the target variable, which is an integer 0 or 1, for genuine and fake note respectively. Note, we haven't performed any of these steps ourselves, the dataset is readily available on kaggle \cite{BNKaggle2018} which we use directly for further analysis in the following Sec. \ref{Sec4AA}. 

\subsection{Elementary analysis on banknote dataset \label{Sec4AA}}
\begin{figure*}[!ht]
\centering
\begin{subfigure}{0.324\linewidth}
\includegraphics[width=\linewidth]{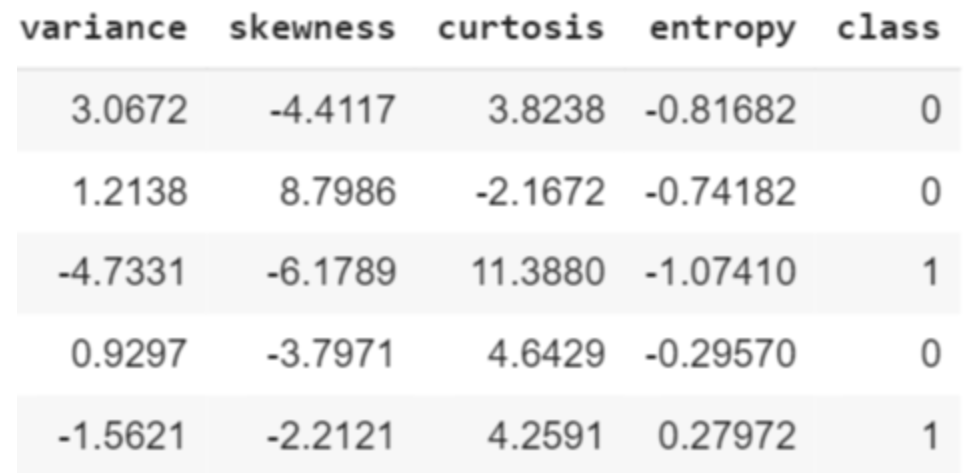} 
\caption{\textbf{Head of banknote dataset.}}
\label{QSVM_Fig4a}
\end{subfigure}
\begin{subfigure}{0.526\linewidth}
\includegraphics[width=\linewidth]{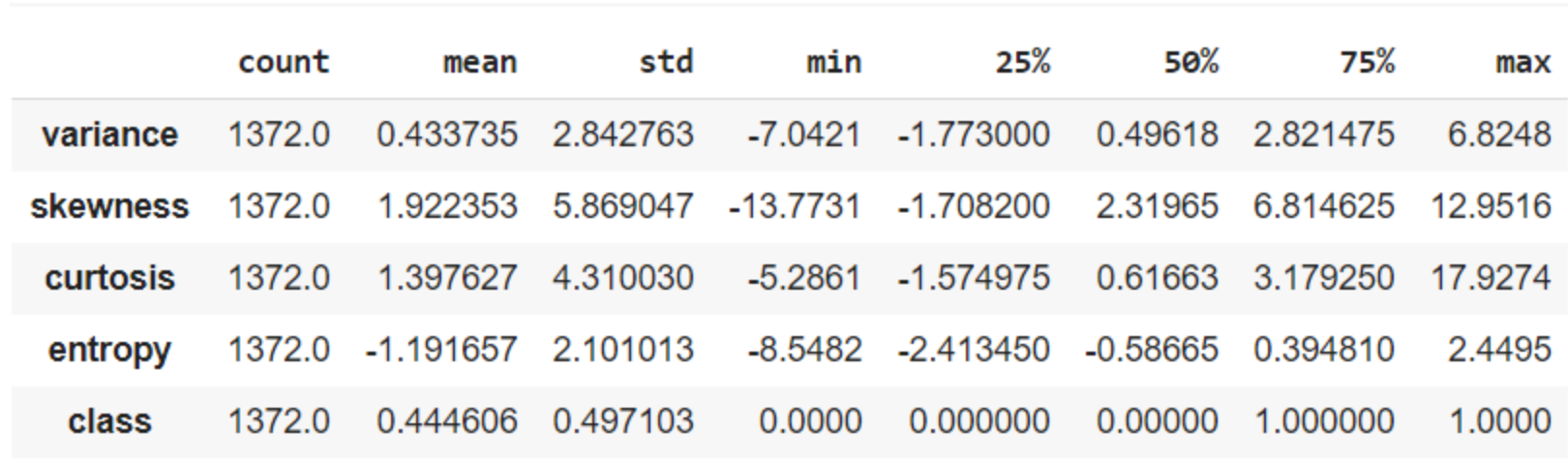} 
\caption{\textbf{Statistics; columns of banknote dataset.}}
\label{QSVM_Fig4b}
\end{subfigure}
\begin{subfigure}{0.90\linewidth}
\includegraphics[width=\linewidth]{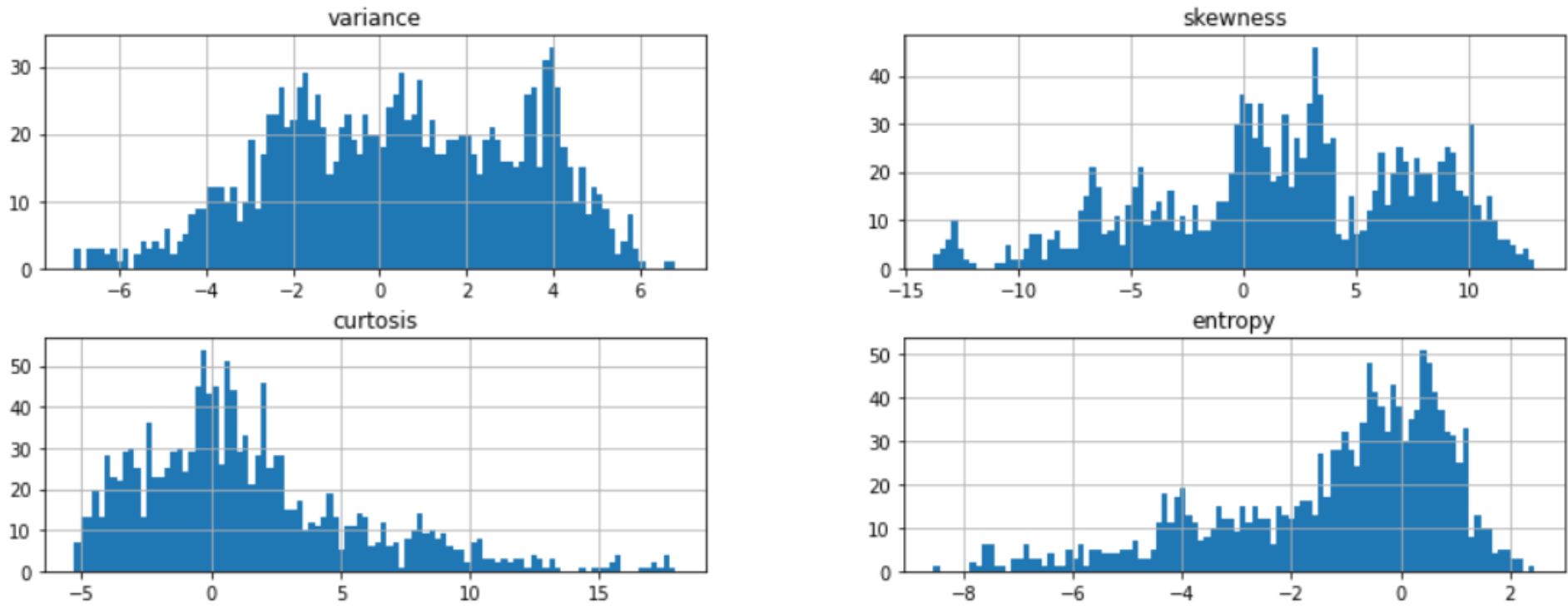} 
\caption{\textbf{Histogram of variance, skewness, kurtosis, entropy.}}
\label{QSVM_Fig4c}
\end{subfigure}
\begin{subfigure}{0.45\linewidth}
\includegraphics[width=\linewidth]{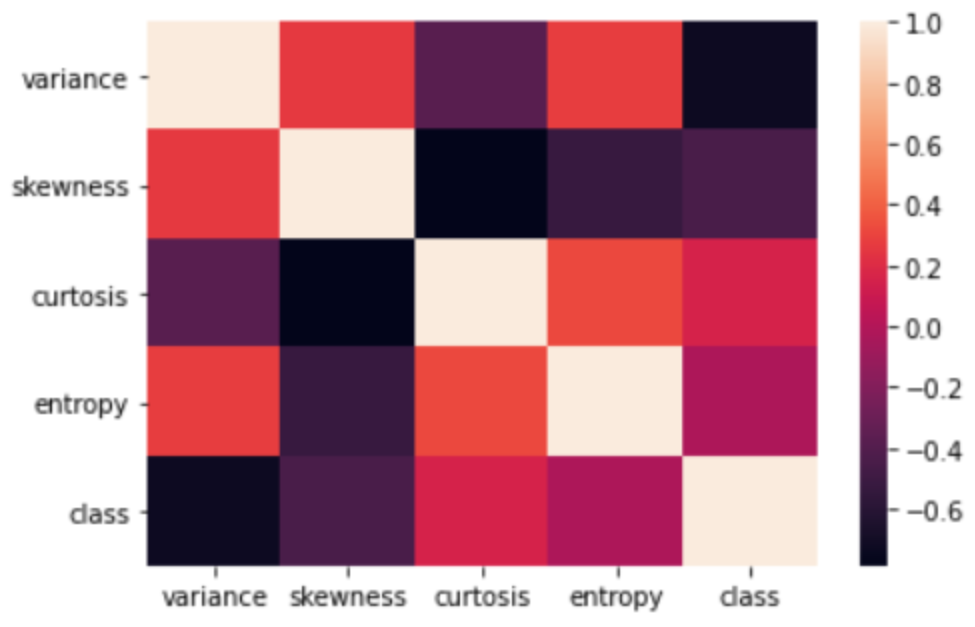} 
\caption{\textbf{Heat map.}}
\label{QSVM_Fig4d}
\end{subfigure}
\begin{subfigure}{0.307\linewidth}
\includegraphics[width=\linewidth]{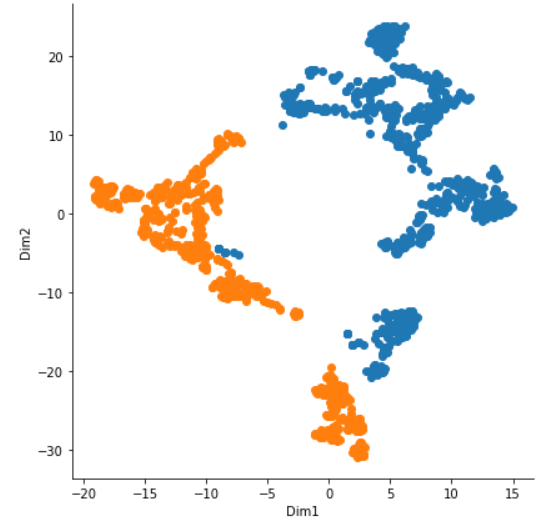}
\caption{\textbf{T-sne plot.}}
\label{QSVM_Fig4e}
\end{subfigure}
\caption{\textbf{Banknote dataset overview: } \textbf{(a)} Here, the head of dataset shows the first five rows. \textbf{(b)} It computes the summary of statistics pertaining to the columns in the banknote dataset, i.e., variance, skewness, kurtosis, entropy, and class. Statistics include count, mean, standard deviation (std), minimum, $25^{th}$ percentile, $50^{th}$ percentile, $75^{th}$ percentile, and maximum. \textbf{(c)} The histograms help visualise the spread of the features of banknote dataset. \textbf{(d)} Here, the heat map represents the correlation between the 4 features: values closer to; 1 represent a positive correlation, -1 a negative correlation, 0 no linear trend. Using the legend in the figure, we can map different colours to different correlation trends among the features. \textbf{(e)} T-sne plot with parameters; perplexity = 100, components=2, iter=10000.}
\label{QSVM_Fig4}
\end{figure*}

Let us now look at elementary machine learning techniques on our dataset, to learn and understand it better. First it is always better to get the summary of statistics for the features in the dataset (Fig. \ref{QSVM_Fig4b}). We observe that out of the 1372 points, 762 belong to class 0 (genuine note) and 610 belong to class 1 (fake note). In Fig. \ref{QSVM_Fig4c} we look at the histogram of variance, skewness, kurtosis, entropy. Histogram is a data representation technique, used to give a visual idea of the probability density function that governs the feature \cite{Histogram-wiki}. Heat map is one of the ways for representing the correlation matrix, and for the banknote dataset, it is shown in Fig. \ref{QSVM_Fig4c}. A correlation matrix denotes the correlation coefficients between variables. A positive correlation indicates a strong dependency, while a negative correlation indicates a strong inverse dependency; a correlation coefficient closer to zero indicates weak dependence \cite{CorrelMatrix-toDS}.

\begin{table}[H]
    \caption{\textbf{Description of features; banknote dataset \cite{ShahaniIJCA2018}}. Here, we describe what each feature represents in our banknote dataset, we also include formulae for variance, skewness and kurtosis for completeness of discussion.}
    \begin{tabular}{|>{\centering}p{2.75cm}|>{\centering}p{5.25cm}|}
        \hline
        Feature & Description \tabularnewline
        \hline
         Variance of wavelet transformed image \\ \vspace{0.25cm} $S^2 = \dfrac{\sum (x_i - \bar{x})^2}{n-1}$ \vspace{0.1cm} & Measure of how each pixel varies from the neighboring pixels (or central pixel), which is used to classify each pixel in different regions \cite{Mathewrg2014}. \tabularnewline
         \hline
         Skewness of wavelet transformed image \\ \vspace{0.25cm} $\tilde{\mu_3} = \dfrac{\sum_{i}^{N} (X_i - \bar{X})^3}{(N-1)*\sigma^3}$ & Measure of the lack of symmetry. Positive skewness; indicates data is skewed towards right, i.e. right tail is long relative to the left tail, and negative skewness is visa versa. Symmetric data like a normal distribution has a skewness near zero \cite{SkewKurt-nist}. \tabularnewline
         \hline
         Kurtosis of wavelet transformed image \\ \vspace{0.25cm} $\mathrm{Kurt} = \dfrac{\mu_4}{\sigma^4}$ & Measure of whether the data are heavy-tailed or light-tailed relative to a normal distribution. Positive kurtosis, i.e heavy-tailed indicates presence of outliers, where as negative kurtosis, i.e. light tailed indicates lack of outliers \cite{SkewKurt-nist}. \tabularnewline 
         \hline
         Entropy of image & 
         Measure of the degree of randomness in the image. It is lower bound for the average coding length (usually measured in bits per pixel) describing amount of information that must be coded by a compression algorithm, thus giving optimum image coding scheme without any loss of information \cite{ThumOpiActa1984}.  
         \tabularnewline
         \hline
    \end{tabular}
    \label{QSVM_table}
\end{table}

\begin{figure}
\centering
\includegraphics[scale=.64]{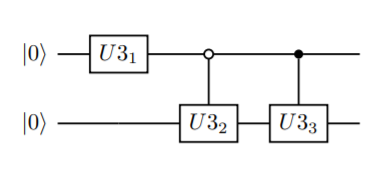}
\caption{\textbf{Encoding circuit for a data of 4 features.} Here, we use a 2-qubit system to encode a dataset with 4 features, it is a direct consequence of steps shown in Eq. \ref{QSVM_Eq6}. For a general 4 feature state $ \ket{\psi_{en}} =  \alpha\ket{00} + \beta\ket{01} + \gamma\ket{10} + \delta\ket{11}$ unitary gate parameters ($U3_1$, $U3_2$, $U3_3$) are shown in Eq. \ref{QSVM_Eq7}.}
\label{QSVM_Fig5}
\end{figure}

\subsubsection{Train and test \label{Sec4A1}}
In machine learning deciding the train and test spilt is important decision, training points are used to train the algorithm and testing points are used as unseen data to estimate algorithm's accuracy. It is common practice in classical ML to spilt the dataset into 80 percent train and 20 percent test, but for our quantum version we identified two main factors which we could consider.  
\begin{itemize}
    \item Firstly, we need as many train samples as possible, so as to better estimate the standard train 1 and standard train 2 for banknote dataset. We see in 6/9 dataset (Fig. \ref{QSVM_Fig2a}) we simply use the standard font for digit 6 and 9, but here we need to define a notion for standard banknote using the training dataset, details for which we explore in Sec. \ref{Sec4B}. 
    \item Secondly, we have to run our QSVM circuit for each of the test data, which would incur computational costs, thus we try to limit the number of test points. In the 6/9 dataset, we had only 8 test samples.
\end{itemize} 

Thus, after trying different percentages for train and test spilt, we think it is the best to have only 28 test samples (which is around 2 \%) and remaining 1344 instances as train samples.

\subsubsection{t-SNE \label{Sec4A2}}
Here, we employ a famous visualisation technique called t-distributed stochastic neighbor embedding (t-SNE) \cite{vanderMaatenJoMLR2008}. We know that our dataset was obtained after digitization of banknote specimen which were either forged or real, but it is possible for the dataset we finally have doesn't naturally represent two clusters. A possible cause could be errors arising at digitization or wavelet transformation step, which may lead to a dataset which represents more or less clusters than expected. Thus we apply we t-SNE which is based on the statistical method of stochastic neighbor embedding \cite{HintonNIP} on our banknote dataset. It is a nonlinear dimensionality reduction technique \cite{NonlinearDRe2007} which helps interpret higher dimensional data, by embedding it in lower dimensional space. Normally a nonlinear dimensionality reduction technique is used to separate data that cannot be simply separated by a straight line \cite{tsne-toDS}. To aid visualisation in t-SNE, lower dimensional space is usually of two or three dimensions. The t-SNE algorithm has two main steps.

\begin{enumerate}
    \item It constructs a probability distribution over pairs of high-dimensional objects by assigning higher probability to similar objects (based on distance) and vice versa. 
    \item In the low-dimensional map, it represents a similar kind of probability distribution, and the Kullback-Leibler (KL) divergence is minimised \cite{tsne-wiki}.
\end{enumerate}

Thus, t-SNE is capable of capturing local structure of the high-dimensional data and reveals global structure such as the presence of clusters \cite{vanderMaatenJoMLR2008}. Using the t-SNE technique, our banknote dataset is visualized in two dimension as shown in Fig. \ref{QSVM_Fig4e}. When run we finally get the best fit parameters as; perplexity = 100, components=2, iter=10000. In Fig. \ref{QSVM_Fig4e} we can see that our dataset indeed has two clusters formed naturally using the t-SNE technique. Draw back with t-SNE is, it is only a visualisation technique, and there is no way to find the values of the clusters centers in Fig. \ref{QSVM_Fig4e}. Even if we try to estimate clusters centers, we directly can't use it, as we would need to retrace it back (from the lower dimension, here 2) to the corresponding feature values (higher dimension, here 4) of the banknote dataset \cite{tsne-stackexch}. Feature values if obtained would have been very useful to standardizing the train 1 and train 2 for banknote dataset, details of which we see in Sec. \ref{Sec4B}.

\subsection{QSVM on banknote dataset}\label{Sec4B}

\subsubsection{Difficulty \label{Sec4B1}}
Two challenges we faced in applying the QSVM algorithm for the bank note dataset is, 
\begin{itemize}
    \item How do we train our QSVM circuit on these 1344 training points?
    \item How do we encode four features of the banknote dataset into the QSVM algorithm?
\end{itemize}

We could recall that in the QSVM general circuit (Fig. \ref{QSVM_Fig1e}) we can feed only one training point pertaining to class 0 (train 1) and one training point pertaining to class 1 (train 2). After splitting the whole dataset of 1372 points into 1344 training points and 28 testing points, we observe that, out of those 1344 training points we have 745 pertaining to class 0 and pertaining to 599 class 1. But then that's the challenge; how do we choice the best possible point, one for class 0 and one for class 1 out these 745 and 599 points respectively? Do we necessarily need to use one of these points or we can use a pre-processing method to best estimate the values of such a point (if it existed)? To understand this better let us first understand what we did in Sec. \ref{Sec3A} for 6/9 case, we notice how we used times new roman (standard font) to standardise the training data to help easily get train 1 and train 2 (Fig. \ref{QSVM_Fig2d}). It is the best possible effort for the 6/9 case, as there it is easy to understand what a standard font would look like, unlike the banknote case where a standard genuine note and standard fake note, do not make sense. In such a scenario, we propose to use classical pre-processing method to help devise such a notation for the banknote case, we understand even if such a point doesn't exist our intention is to estimate best possible train 1 and train 2 for our banknote dataset. In the later sections of averages (Sec. \ref{Sec4B3A}) and k-means (Sec. \ref{Sec4B3K}) we propose two classical pre-processing methods we could use. Aim of such a classical pre-processing method is to obtain values of the 4 features (variance, skewness, kurtosis, entropy), one corresponding to standard training sample train 1 (class 0) and standard training sample train 2 (class 1). But, to finally get the train 1 and train 2 circuit we will need a four feature encoding procedure which will take these values as input, and generate the corresponding training circuits. We propose one such encoding procedure using a 2-qubit system in the coming section \ref{Sec4B2}.

\subsubsection{Encoding of Parameters \label{Sec4B2}}

In the Sec. \ref{Sec2} we recall that the training data oracle (Fig. \ref{QSVM_Fig1c}) and unitary operator $U_{x0}$ (Fig. \ref{QSVM_Fig1b}) could encode only one feature at first, i.e., the theta value of the $R_y$ gates. Which we later see in the 6/9 case (Sec. \ref{Sec3}); it could easily be used to extend for 2 features (a vector $\vec{x} = (\alpha,\beta)$) using Eq. \eqref{QSVM_Eq5}. To apply QSVM algorithm to our banknote dataset, we need an encoding procedure, to encode the values of the 4 features; which were variance, skewness, kurtosis, and entropy. Let us assume the normalised four features as, $\alpha$, $\beta$, $\gamma$, and $\delta$. In quantum computing it is standard practice to use normalized values, which implies, $\alpha^2$ + $\beta^2$ + $\gamma^2$ + $\delta^2$ = 1. The state encoding the four features would be $ \ket{\psi_{en}} =  \alpha\ket{00} + \beta\ket{01} + \gamma\ket{10} + \delta\ket{11}$. It is difficult to predict the quantum circuit that can convert $\ket{00}$ to $\ket{\psi_{en}}$ as there could be large number of possibilities for states, due to large number of unitaries. However, it is easier to find the circuit that can convert $\ket{\psi_{en}}$ to $\ket{00}$, which then can be used to obtain $\ket{00}$ to $\ket{\psi_{en}}$ as quantum operations are reversible. The $\ket{\psi_{en}}$ state can also be written as shown in Eq. \ref{QSVM_Eq6}. Following steps in Eq. \ref{QSVM_Eq6}, are series of anti-control unitary and control unitary operations to convert $\ket{\psi_{en}}$ to $\ket{00}$, which lead to quantum circuit in Fig. \ref{QSVM_Fig5}. 

\begin{figure*}[!ht]
\centering
\begin{subfigure}{0.48\linewidth}
\includegraphics[width=\linewidth]{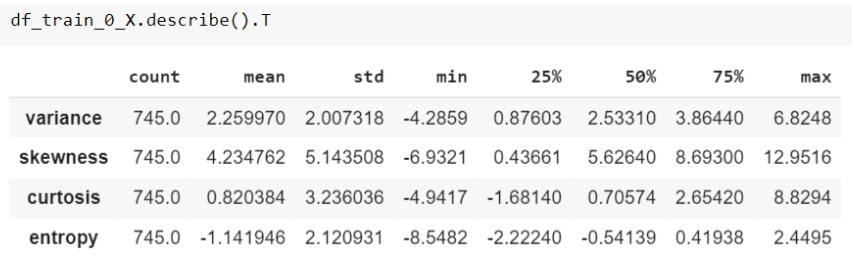}
\caption{\textbf{Statistics; Class 0 split.}}
\label{QSVM_Fig7a}
\end{subfigure}
\begin{subfigure}{0.505\linewidth}
\includegraphics[width=\linewidth]{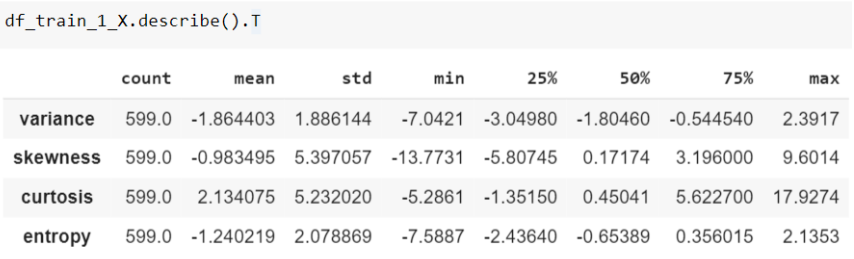} 
\caption{\textbf{Statistics; Class 1 split.}}
\label{QSVM_Fig7b}
\end{subfigure}
\begin{subfigure}{0.495\linewidth}
\includegraphics[width=\linewidth]{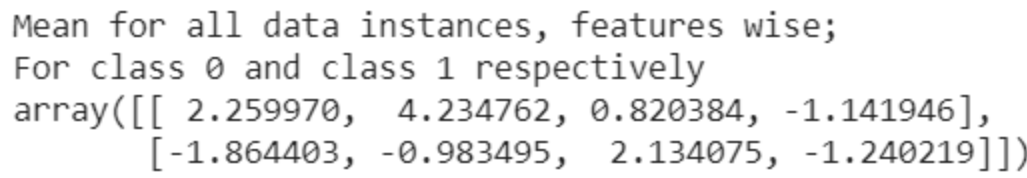} 
\caption{\textbf{Averages; cluster centers} }
\label{QSVM_Fig7c}
\end{subfigure}
\begin{subfigure}{0.345\linewidth}
\includegraphics[width=\linewidth]{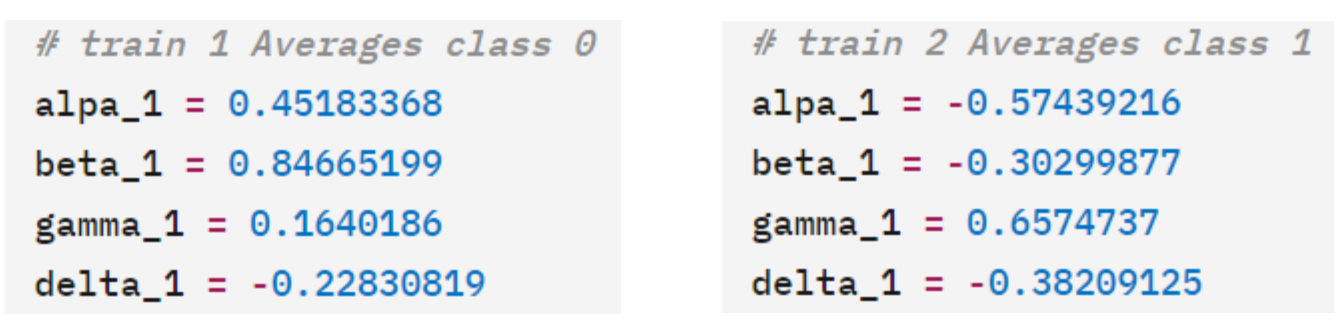} 
\caption{\textbf{Averages; normalised cluster centers} }
\label{QSVM_Fig7d}
\end{subfigure}
\begin{subfigure}{0.213\linewidth}
\includegraphics[width=\linewidth]{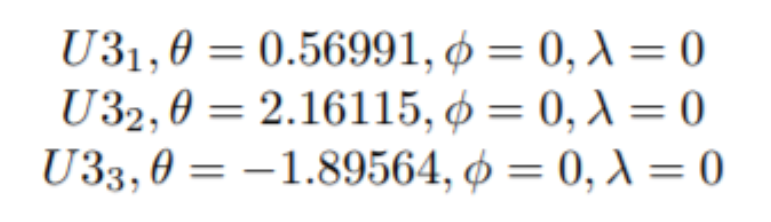} 
\caption{\textbf{Train 1 parameters}}
\label{QSVM_Fig7e}
\end{subfigure}
\begin{subfigure}{0.19\linewidth}
\includegraphics[width=\linewidth]{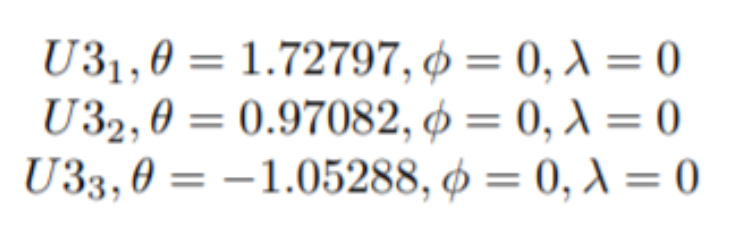}
\caption{\textbf{Train 2 parameters}}
\label{QSVM_Fig7f}
\end{subfigure}
\begin{subfigure}{0.26\linewidth}
\includegraphics[width=\linewidth]{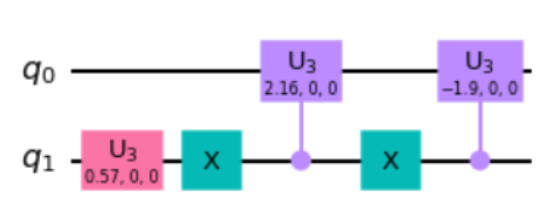} 
\caption{\textbf{Train 1 circuit}}
\label{QSVM_Fig7g}
\end{subfigure}
\begin{subfigure}{0.31\linewidth}
\includegraphics[width=\linewidth]{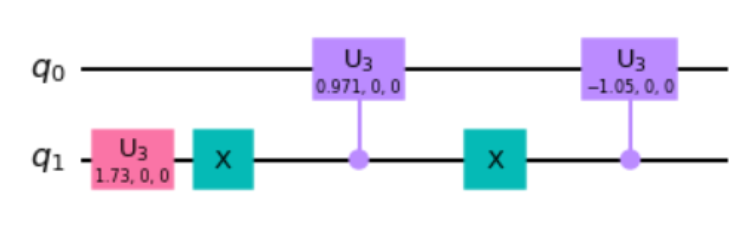}
\caption{\textbf{Train 2 circuit}}
\label{QSVM_Fig7h}
\end{subfigure}
\caption{{\textbf{Averages approach; banknote dataset: } \textbf{(a)} and \textbf{(b)}: It computes the summary of statistics pertaining to the columns i.e., variance, skewness, kurtosis, and entropy for class 0 and class 1 respectively. Statistics include count, mean, standard deviation (std), minimum, $25^{th}$ percentile, $50^{th}$ percentile, $75^{th}$ percentile, and maximum. \textbf{(c)} Values represent mean of variance, skewness, kurtosis, and entropy for class 0 spilt and class 1 spilt respectively, which is an immediate results from Figs. \ref{QSVM_Fig7a} and \ref{QSVM_Fig7b}. \textbf{(d)} Cluster centers from Fig. \ref{QSVM_Fig7c} needs to be normalised before it can be used in our encoding procedure proposed in Sec. \ref{Sec4B2}. Here, train 1 represents normalised cluster centers for class 0 and train 2 represents normalised cluster centers for class 1. \textbf{(e)} and \textbf{(f)}: Represents the encoding parameters for train 1 and train 2 in the averages approach respectively. \textbf{(g)} and \textbf{(h)}: Represents the encoding circuit for train 1 and train 2 respectively, which follows directly from Figs. \ref{QSVM_Fig7e} and \ref{QSVM_Fig7f}, using the general encoding circuit from Fig. \ref{QSVM_Fig5}.}}
\label{QSVM_Fig7}
\end{figure*}

\begin{eqnarray}
\ket{\psi_{en}} &=& \alpha\ket{00} + \beta\ket{01} + \gamma\ket{10} + \delta\ket{11}  \nonumber\\
&=& \sqrt{|\alpha|^2 + |\beta|^2}\ket{0}(\dfrac{\alpha\ket{0} + \beta\ket{1}}{\sqrt{|\alpha|^2 + |\beta|^2}}) \nonumber\\ 
&+&\sqrt{|\gamma|^2 + |\delta|^2}\ket{1}(\dfrac{\gamma\ket{0} + \delta\ket{1}}{\sqrt{|\gamma|^2 + |\delta|^2}}) \nonumber\\
&=&\sqrt{|\alpha|^2 + |\beta|^2}\ket{0}(\ket{0}) + \sqrt{|\gamma|^2 + |\delta|^2}\ket{1}(\ket{0})\nonumber\\    
&=& (\sqrt{|\alpha|^2 + |\beta|^2}\ket{0} + \sqrt{|\gamma|^2 + |\delta|^2}\ket{1})\ket{0} \nonumber\\
&=& \ket{00}
\label{QSVM_Eq6}
\end{eqnarray}

We know a general $U3$ gate has parameters,

\begin{eqnarray}
U(\theta, \phi, \lambda)=\left[\begin{matrix}\cos{\frac{\theta}{2}} &-e^{i \lambda}\sin{\frac{\theta}{2}}\\e^{i \phi}\sin{\frac{\theta}{2}} & e^{i(\phi+\lambda)}\cos{\frac{\theta}{2}}\end{matrix}\right] 
\label{qsvm_Eq7b}
\end{eqnarray}

Parameters of the gates $U3_1$, $U3_2$, $U3_3$ in the encoding circuit (Fig. \ref{QSVM_Fig5}) are as shown in Eq. \eqref{QSVM_Eq7}.

\begin{center}
\begin{eqnarray}
U3_1, \theta &=& 2\tan^{-1}(\dfrac{\sqrt{|\gamma|^2 + |\delta|^2}}{\sqrt{|\alpha|^2 + |\beta|^2}}) , \phi = 0, \lambda = 0 \nonumber \\
U3_2, \theta &=& 2\tan^{-1}(\dfrac{\beta}{\alpha}), \phi = 0, \lambda = 0 \nonumber \\
U3_3, \theta &=& 2\tan^{-1}(\dfrac{\gamma}{\delta}), \phi = 0, \lambda = 0
\label{QSVM_Eq7}
\end{eqnarray}
\end{center}

\subsubsection{Averages}\label{Sec4B3A}

Here we propose a classical pre-processing method, which we call averages. In this approach of classical pre-processing, we first split the training dataset containing of 1344 rows into class 0 and class 1. Then, we find statistics pertaining to features for respective class, which can be seen in Fig. \ref{QSVM_Fig7a} for class 0 and Fig. \ref{QSVM_Fig7b} for class 1. Our training dataset has 745 points of class 0 and 599 points of class 1. In our approach of averages from tables (Fig. \ref{QSVM_Fig7a} and Fig. \ref{QSVM_Fig7b}) we take mean for all features for class 0 and class 1 respectively, which can be found as cluster centers in Fig. \ref{QSVM_Fig7c}. To use the proposed encoding (Sec. \ref{Sec4B2} ); for encoding 4 features into 2-qubit system, we need to normalise cluster centers. Fig. \ref{QSVM_Fig7d} shows the normalised cluster centers, where train 1 represents class 0 and train 2 represents class 1. Fig. \ref{QSVM_Fig7e} represents the unitary gate parameters for train 1 which follows directly from the encoding method using the values from Fig. \ref{QSVM_Fig7d}. Similarly, Fig. \ref{QSVM_Fig7f} represents the unitary gate parameters for train 2. Figs. \ref{QSVM_Fig7g} and \ref{QSVM_Fig7h} show the corresponding train 1 and train 2 circuits.


\subsubsection{K means}\label{Sec4B3K}
\begin{figure*}[!ht]
\centering
\begin{subfigure}{0.48\linewidth}
\includegraphics[width=\linewidth]{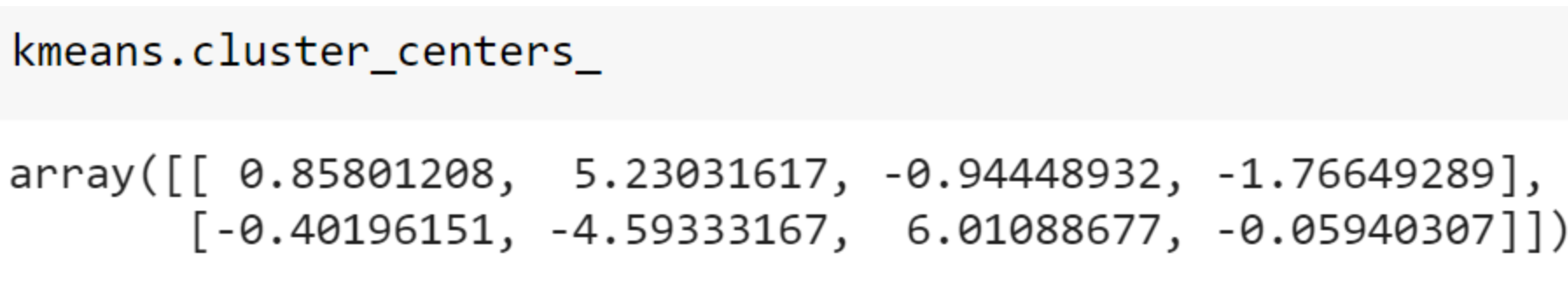} 
\caption{\textbf{K Means Centers} }
\label{QSVM_Fig8a}
\end{subfigure}
\begin{subfigure}{0.48\linewidth}
\includegraphics[width=\linewidth]{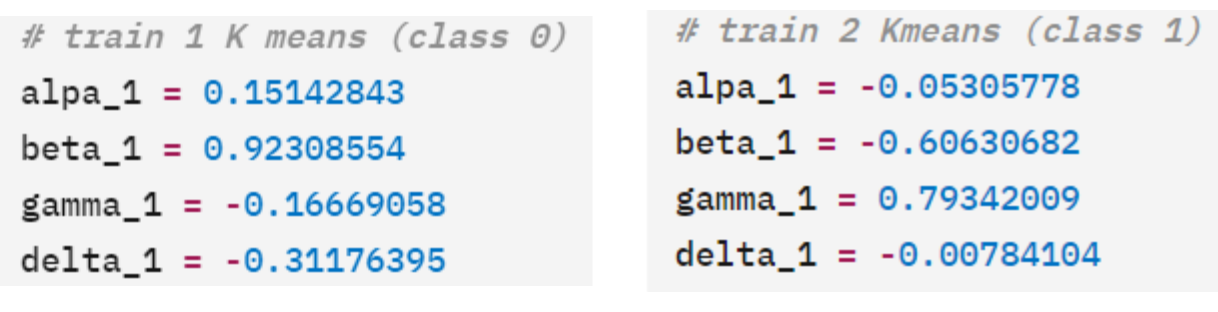} 
\caption{\textbf{K Means Normalise} }
\label{QSVM_Fig8b}
\end{subfigure}
\begin{subfigure}{0.193\linewidth}
\includegraphics[width=\linewidth]{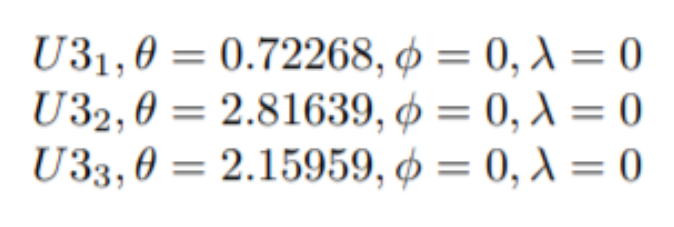}
\caption{\textbf{Train 1 parameters}}
\label{QSVM_Fig8c}
\end{subfigure}
\begin{subfigure}{0.21\linewidth}
\includegraphics[width=\linewidth]{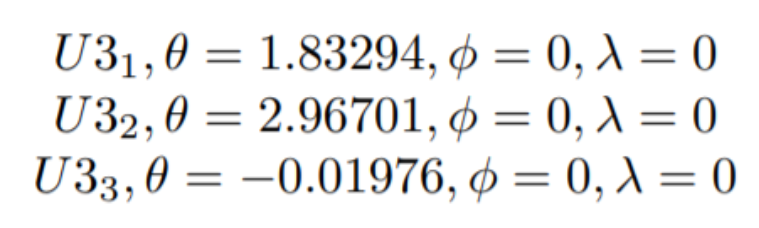}
\caption{\textbf{Train 2 parameters}}
\label{QSVM_Fig8d}
\end{subfigure}
\begin{subfigure}{0.277\linewidth}
\includegraphics[width=\linewidth]{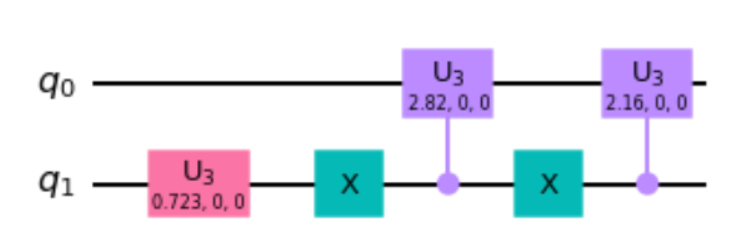} 
\caption{\textbf{Train 1 circuit}}
\label{QSVM_Fig8e}
\end{subfigure}
\begin{subfigure}{0.29\linewidth}
\includegraphics[width=\linewidth]{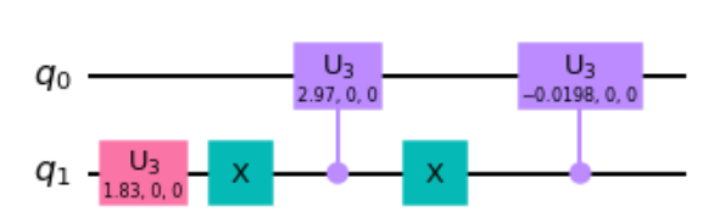}
\caption{\textbf{Train 2 circuit}}
\label{QSVM_Fig8f}
\end{subfigure}
\caption{{\textbf{K-means approach; banknote dataset: } \textbf{(a)} Represents the cluster centers obtained by the k-means algorithm in the order; variance, skewness, kurtosis, and entropy for class 0 and class 1 respectively. \textbf{(b)}: K-means centers from Fig. \ref{QSVM_Fig8a} needs to be normalised before it can be used in our encoding procedure proposed in Sec. \ref{Sec4B2}. Here, train 1 represents normalised cluster centers for class 0 and train 2 represents normalised cluster centers for class 1. \textbf{(c)} and \textbf{(d)}: Represents the encoding parameters for train 1 (class 0) and train 2 (class 1) in the k-means approach respectively. \textbf{(e)} and \textbf{(f)}: Represents the encoding circuit for train 1 (class 0) and train 2 (class 1) respectively, which follows directly from Figs. \ref{QSVM_Fig8c} and \ref{QSVM_Fig8d}, using the general encoding circuit from Fig. \ref{QSVM_Fig5}.}}
\label{QSVM_Fig8}
\end{figure*}

Here, we propose a second classical pre-processing method, which we call k-means. In this approach, we apply the classical K-means algorithm \cite{Lloydieee1982} on our training instances and then find k-means cluster centers, which is shown in Fig. \ref{QSVM_Fig8a}. Again, to use the encoding procedure proposed in Sec. \ref{Sec4B2}, we need to normalise these k-means cluster centers. Fig. \ref{QSVM_Fig8b} shows the normalised k-means cluster centers, where train 1 represents class 0 and train 2 represents class 1. Fig. \ref{QSVM_Fig8c} represents the unitary gate parameters for train 1 which follows directly from the encoding method using the values from Fig. \ref{QSVM_Fig8b}. Similarly, Fig. \ref{QSVM_Fig8d} represents the unitary gate parameters for train 2. Figs. \ref{QSVM_Fig8e} and \ref{QSVM_Fig8f} show the corresponding train 1 and train 2 circuits.


\begin{figure*}
\centering
\includegraphics[width=\textwidth]{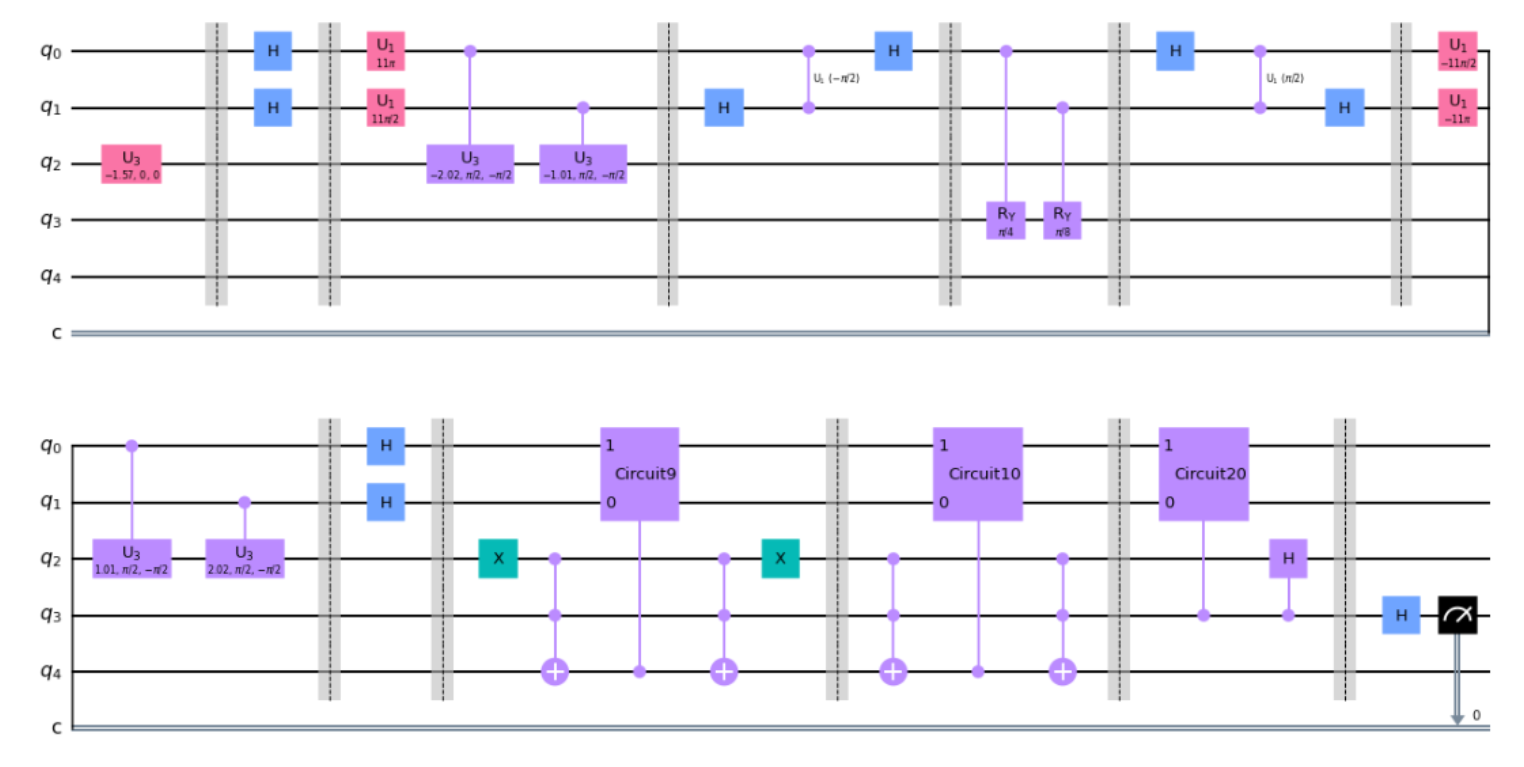}
\caption{\textbf{QSVM circuit for banknote :} Here, the parameters of the general circuit from Fig. \ref{QSVM_Fig1e}; $R_y1$ $R_y2$ and then query state $R_y0$ consists 2 qubits using the proposed encoding from Fig. \ref{QSVM_Fig5}. It is to be noted that here we use two methods, averages \ref{Sec4B3A} and K means \ref{Sec4B3A} to find the parameters of the circuit 9, 10 and 20 which refer to $R_y1$, $R_y2$ and $R_y0$ respectively.}
\label{QSVM_Fig69}
\end{figure*}

\subsubsection{QSVM Circuit for BN \label{Sec4B4}}

To be able to apply QSVM algorithm on banknote dataset, according to the QSVM general circuit (Fig. \ref{QSVM_Fig1e}) we need the F matrix,   
$R{y}^{\theta_{1}}$, $R{y}^{\theta_{2}}$ and $R{y}^{-\theta_{0}}$. Parameters $R{y}^{\theta_{1}}$, $R{y}^{\theta_{2}}$ represent the training vectors train 1 and train 2 respectively, which for our banknote case we obtain from the train 1 and train 2 circuits of the classical pre-processing methods as discussed in Sec. \ref{Sec4B3A} and Sec. \ref{Sec4B3A}. The train 1 and train 2 circuits for each case of classical pre-processing would respectively replace the circuit 9 and circuit 10 of the Fig. \ref{QSVM_Fig69}. $R{y}^{-\theta_{0}}$ represents the testing vector which would vary for the 28 tests points. The 28 tests points are obtained after the train-test split of the dataset. Similar to steps discussed in Sec. \ref{Sec4B3A} and Sec. \ref{Sec4B3K} to obtain train 1 and train 2 circuits, we obtain the test 0 circuits by first normalising each test point and then use the encoding procedure (Sec. \ref{Sec4B2}). We haven't explicitly shown test 0 circuits, but the idea can be easily understood from Fig. \ref{QSVM_Fig69}, where we use the circuit 20 as placeholder for test 0 circuits. After having obtained the train 1 and train 2 points from the classical pre-processing, the F matrix is just inner product of them. Thus we now have all the parameters to apply the QSVM algorithm to banknote dataset.

\textbf{Results for banknote case:}
\begin{enumerate}
    \item \textbf{Averages:} 
    \begin{itemize}
        \item IBM qasm simulator, for 8192 shots, correct results obtained for 18/28 test points, i.e., accuracy - \textbf{64.3\%}.
        \item IBM belem, for 8192 shots, correct results obtained for 18/28 test points, i.e., accuracy - \textbf{64.3\%}.
    \end{itemize}
    \item \textbf{K-means:}  
        \begin{itemize}
        \item IBM qasm simulator, for 8192 shots, correct results obtained for 19/28 test points, i.e., accuracy - \textbf{67.9\%}.
        \item IBM belem, for 8192 shots, correct results obtained for 16/28 test points, i.e., accuracy - \textbf{57.1\%}.
    \end{itemize}
\end{enumerate}

These results, just like the 6/9 case could be explained by formation of erroneous solution.

Specification of quantum system used in this section.
\begin{enumerate}
    \item IBM qasm simulator: 32 qubits. Simulator type: General, context-aware.
    \item IBM belem specifications: 5 qubits; 16 quantum volume; 2.5K CLOPS; Avg. T1: 93.81 $\mu$s; Avg. T2: 102.93 $\mu$s.
\end{enumerate}

\section{Our proposed method \label{Sec5}}
Here we propose a new method inspired by the following works \cite{HavlicekNat2019,SchuldPRL2019,SchuldPRL2020,HeredgearXiv2021} which we will apply to both; 6/9  and banknote dataset. Here is our algorithm. 

For the binary classification problem, let us consider the testing vector, as $\ket{\psi}_0$ and the two training vectors $\ket{\psi}_1$ and $\ket{\psi}_2$ representing class 0 and class 1 respectively. To classify the testing vector $\ket{\psi}_0$, we compare it with the two training vectors $\ket{\psi}_1$ and $\ket{\psi}_2$. Comparison is done by taking the inner products; one between $\ket{\psi}_1$ and $\ket{\psi}_0$ and second between $\ket{\psi}_2$ and $\ket{\psi}_0$. Class of the training vector with which the inner product has higher value, is assigned to the testing vector $\ket{\psi}_0$. First, we define the quantum states using the unitaries $U_1$, $U_2$ and $U_0$ such that $\ket{\psi}_1=U_1\ket{00...0}$, $\ket{\psi}_2=U_2\ket{00...0}$ and $\ket{\psi}_0=U_0\ket{00...0}$. Next, let us look at the following terms, $\bra{\psi_1}\ket{\psi_0}=\bra{00..0}{U_1}^{\dagger}U_0\ket{00..0}=\bra{00..0}A_{10}\ket{00..0}$ where $A_{10}={U_1}^{\dagger}U_0$. $\bra{\psi_2}\ket{\psi_0}=\bra{00..0}{U_2}^{\dagger}U_0\ket{00..0}=\bra{00..0}A_{20}\ket{00..0}$ where $A_{20}={U_2}^{\dagger}U_0$. 
We observe that to calculate the inner products $\bra{\psi_1}\ket{\psi_0}$ and $\bra{\psi_2}\ket{\psi_0}$, we need quantum circuits corresponding to $A_{10}$ and $A_{20}$. These circuits can be easily be obtained by encoding the corresponding classical data into the unitaries $U_1$, $U_2$ and $U_0$. $A_{10}$ and $A_{20}$ circuits initialized over state $\ket{00..0}$ would result in a superposition of the terms as, $A_{10}\ket{00..0}=a_{10}\ket{00..0}+b_{10}\ket{00..1}+...$ and $A_{20}\ket{00..0}=a_{20}\ket{00..0}+b_{20}\ket{00..1}+...$. Taking the inner products with $\bra{00..0}$, would result in $a_{10}$, $a_{20}$, which is also the amplitude of $\ket{00..0}$ for $A_{10}$ and $A_{20}$ respectively. Thus the probability of measuring the $A_{10}$ and $A_{20}$ quantum circuits in $\ket{00..0}$ should be equal to the $|\bra{\psi}_1\ket{\psi}_0|^2$ and $|\bra{\psi}_2\ket{\psi}_0|^2$ respectively. Hence, in-order to find the inner products $\bra{\psi}_1\ket{\psi}_0$ and $\bra{\psi}_2\ket{\psi}_0$ we need to create the quantum circuits for $A_{10}$, $A_{20}$ and measure them.

\begin{figure}
\centering
\begin{subfigure}{0.48\linewidth}
\includegraphics[width=\linewidth]{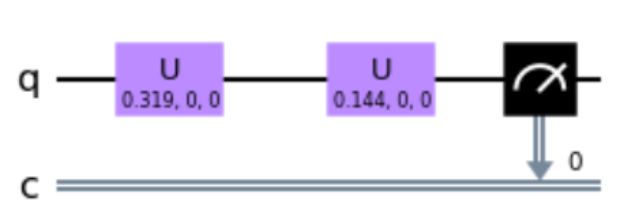} 
\caption{\textbf{$A_{10}$ quantum circuit}}
\label{QSVM_Fig9a}
\end{subfigure}\hfill
\begin{subfigure}{0.48\linewidth}
\includegraphics[width=\linewidth]{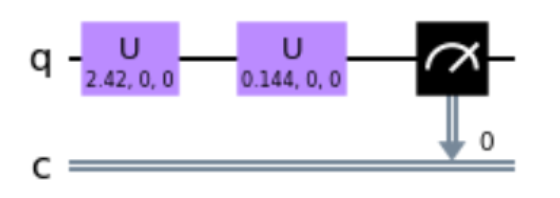} 
\caption{\textbf{$A_{20}$ quantum circuit}}
\label{QSVM_Fig9b}
\end{subfigure}\hfill
\begin{subfigure}{0.99\linewidth}
\includegraphics[width=\linewidth]{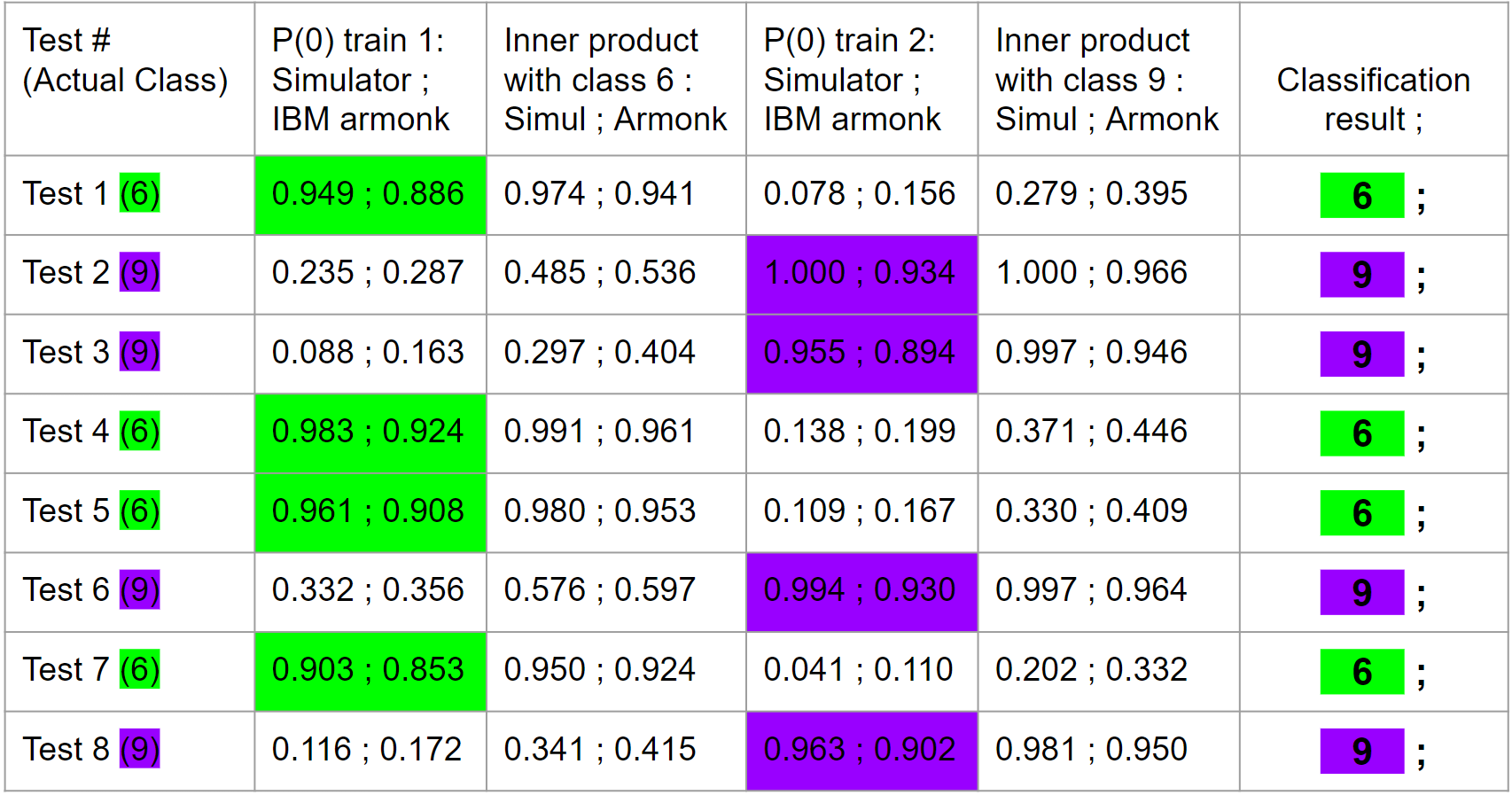}
\caption{\textbf{Results; run on IBM qasm simulator and IBM armonk.}}
\label{QSVM_Fig9c}
\end{subfigure}
\caption{ \textbf{Proposed method; 6/9 data-set} \textbf{(a)} Shows the circuit that calculates the inner product between test 1 and train 1 (class 6). \textbf{(b)} Shows the circuit that calculates the inner product between test 1 and train 2 (class 9). \textbf{(c)} Here we show the results obtained for the 8 test points on IBM qasm simulator and IBM armonk. We tabulate and compare for all 8 test points; their inner products with class 6 and class 9. Classification of each test point is done based on which inner product had higher values.}
\label{QSVM_Fig9}
\end{figure}

For implementation of proposed method on the 6/9 case, we need to first encode our data into the quantum states $\ket{\psi}_1$, $\ket{\psi}_2$ and $\ket{\psi}_0$. We recall in the 6/9 case we have only 2 features, so the Eq. \eqref{QSVM_Eq5}, could be used here to we find the unitaries $U_1$, $U_2$ and $U_0$ corresponding to the quantum states $\ket{\psi}_1$, $\ket{\psi}_2$ and $\ket{\psi}_0$. To get the parameters of the U3 gates ( $U(\theta, \phi, \lambda)$ refer Eq. \eqref{qsvm_Eq7b}) corresponding to the respective unitaries, theta values table from the QSVM case Fig. \ref{QSVM_Fig2d} (Sec. \ref{Sec2}) could be directly used. Now we can easily generate the corresponding $A_{10}$ circuit and $A_{20}$ circuits for all the 8 test points. Fig. \ref{QSVM_Fig9a} and Fig. \ref{QSVM_Fig9b} shows the $A_{10}$ circuit and $A_{20}$ circuit for test 1 respectively. We find the $A_{10}$ and $A_{20}$ circuits for all 8 test points and summarise the results run on IBM quantum simulator and IBM real chip as shown in Fig. \ref{QSVM_Fig9c}.

\begin{figure*}[!ht]
\centering
\begin{subfigure}{0.49\linewidth}
\includegraphics[width=\linewidth]{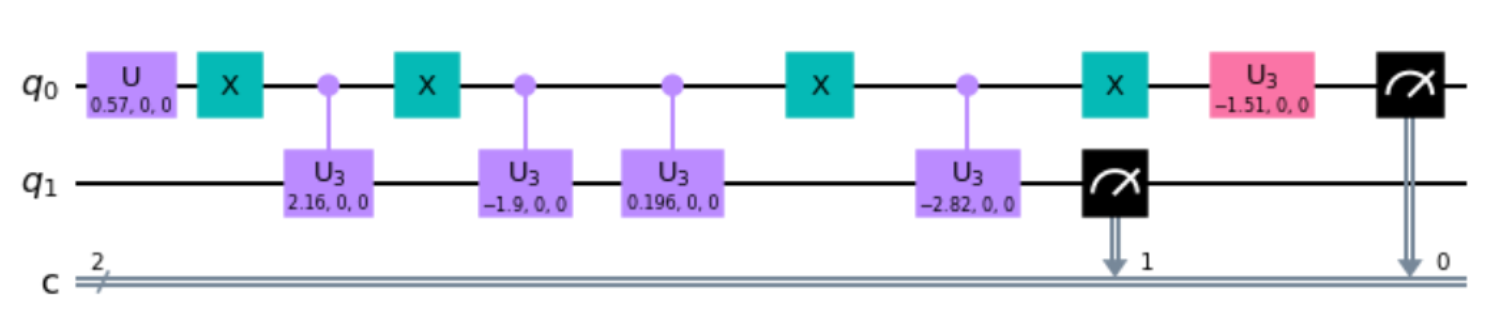}
\caption{\textbf{Averages; $A_{10}$ quantum circuit}}
\label{QSVM_Fig10a}
\end{subfigure}
\begin{subfigure}{0.49\linewidth}
\includegraphics[width=\linewidth]{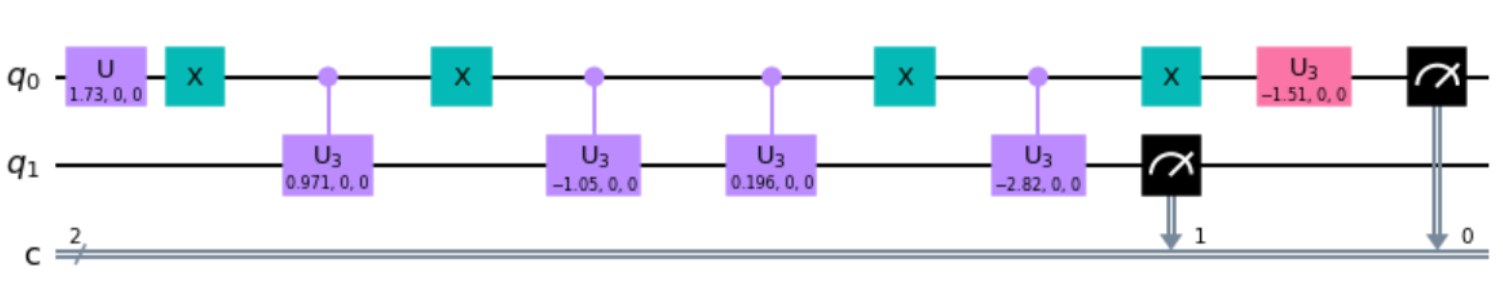}
\caption{\textbf{Averages; $A_{20}$ quantum circuit}}
\label{QSVM_Fig10b}
\end{subfigure}
\centering
\begin{subfigure}{0.49\linewidth}
\includegraphics[width=\linewidth]{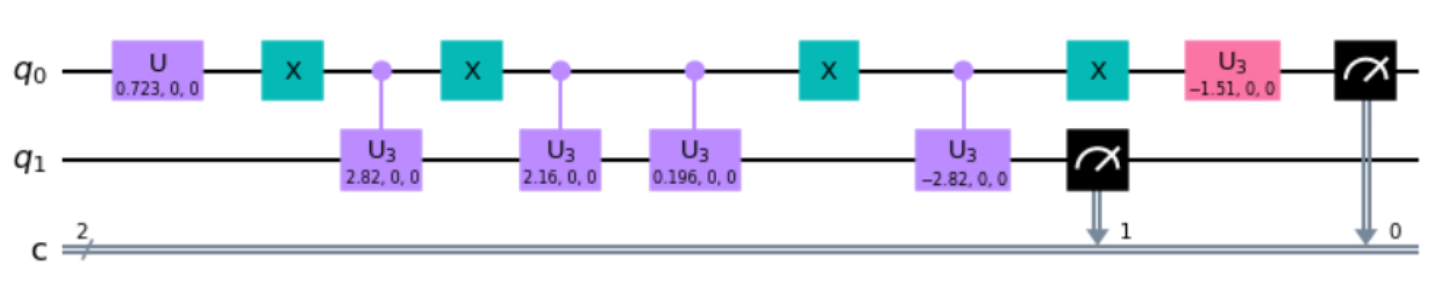} 
\caption{\textbf{K-means; $A_{10}$ quantum circuit}}
\label{QSVM_Fig10c}
\end{subfigure}
\begin{subfigure}{0.49\linewidth}
\includegraphics[width=\linewidth]{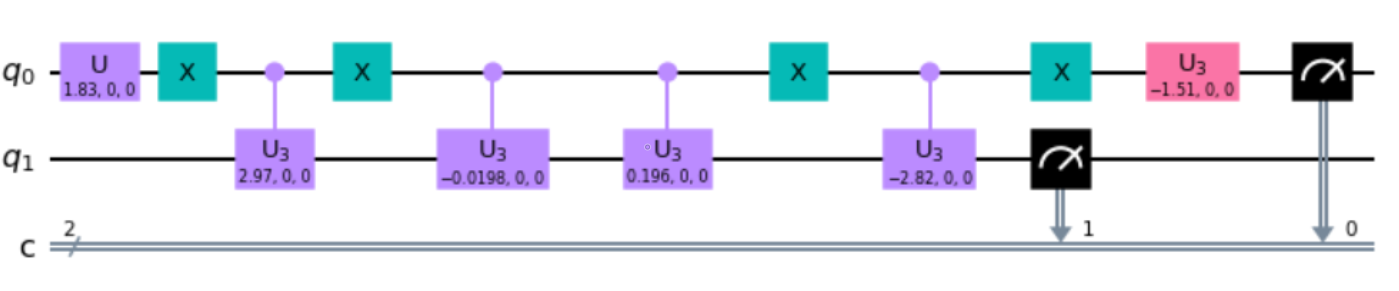}
\caption{\textbf{K-means; $A_{20}$ quantum circuit}}
\label{QSVM_Fig10d}
\end{subfigure}
\begin{subfigure}{0.69\linewidth}
\includegraphics[width=\linewidth]{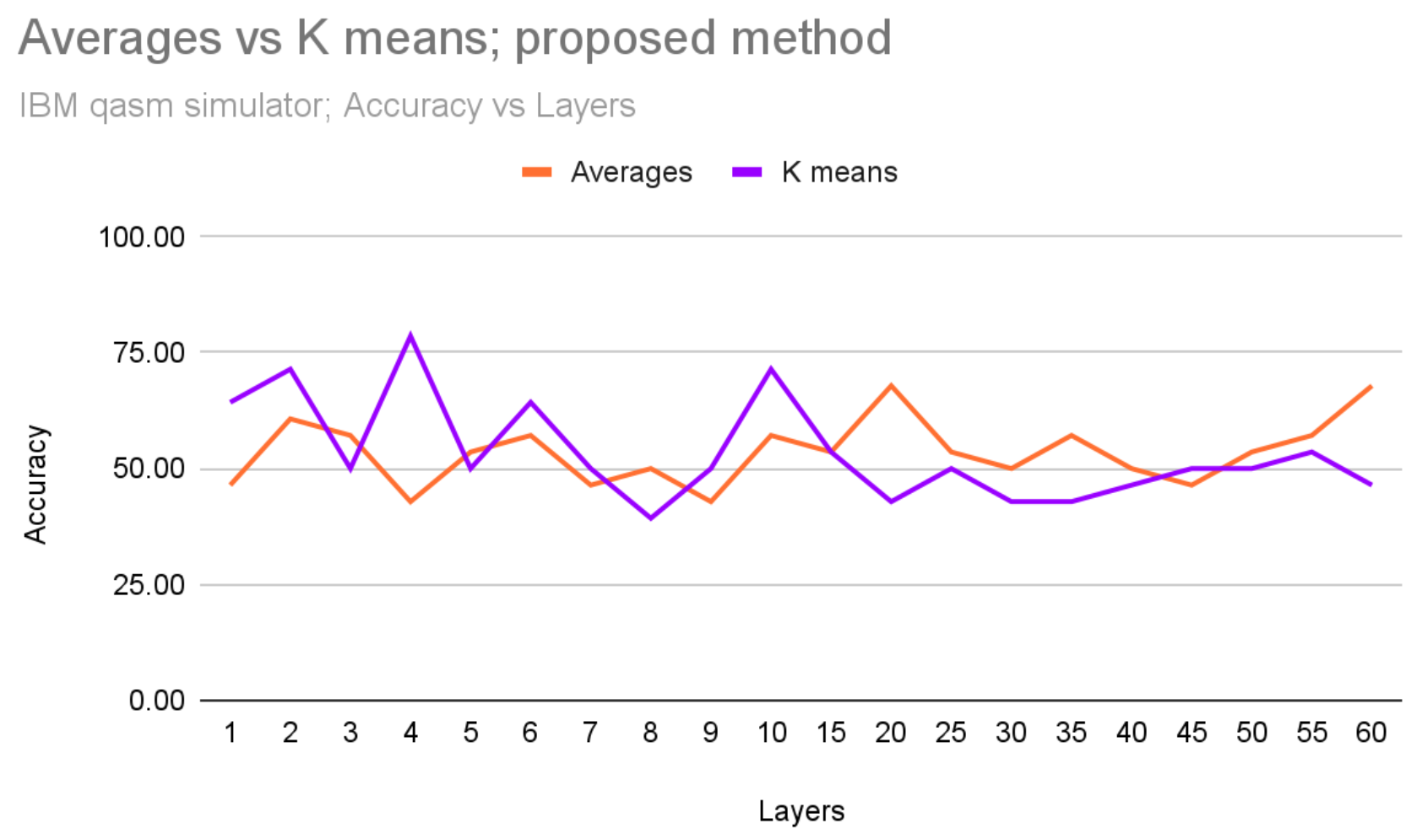}
\caption{\textbf{Results; run on IBM qasm simulator.}}
\label{QSVM_Fig10e}
\end{subfigure}
\begin{subfigure}{0.69\linewidth}
\includegraphics[width=\linewidth]{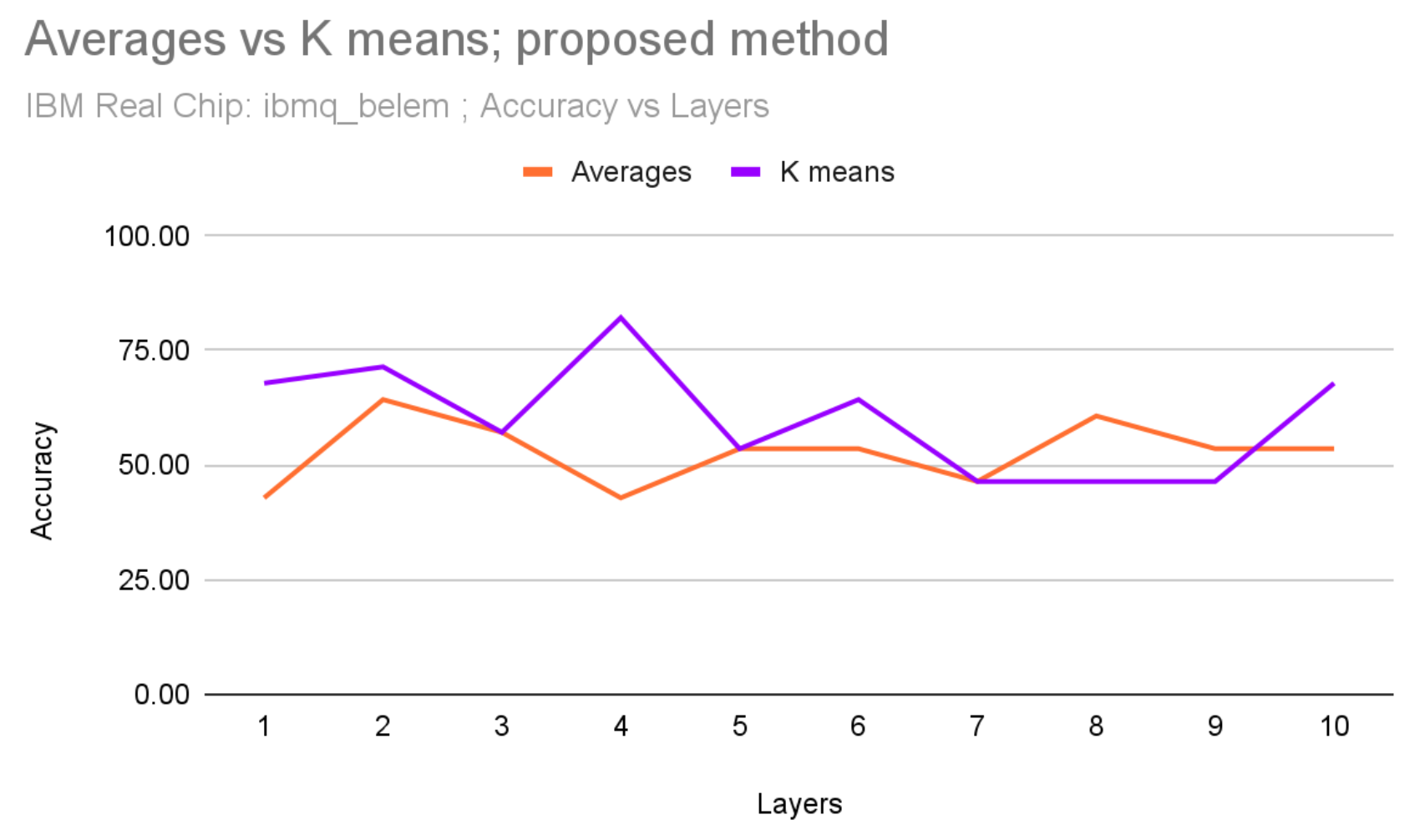}
\caption{\textbf{Results; run on IBM belem.}}
\label{QSVM_Fig10f}
\end{subfigure}
\caption{{ \textbf{Proposed method; banknote data-set} \textbf{(a)} and  \textbf{(b)}: shows the $A_{10}$ and $A_{20}$ quantum circuit for the averages method. $A_{10}$ shows the circuit representing the inner product  between test 1 and train 1 (class 0). Similarly $A_{20}$ represents between test 1 and train 2 (class 1). \textbf{(c)} and \textbf{(d)}: shows the $A_{10}$ and $A_{20}$ quantum circuit for the k-means method. $A_{10}$ shows the circuit representing the inner product  between test 1 and train 1 (class 0). Similarly $A_{20}$ represents between test 1 and train 2 (class 1). \textbf{(e)} Shows the comparison between averages and k-means for banknote data-set; for layers 1 to 60 on IBM qasm simulator. \textbf{(f)} Shows the comparison between averages and k-means for banknote data-set; for layers 1 to 10 on IBM belem.}}
\label{QSVM_Fig10}
\end{figure*}

For implementation of proposed method on the banknote authentication dataset, we need to encode our data into the quantum states corresponding $\ket{\psi}_1$, $\ket{\psi}_2$ and $\ket{\psi}_0$ where $\ket{\psi}_1$ and $\ket{\psi}_2$ correspond to training vector for class 0 and class 1 respectively. We recall, in the banknote dataset out total 1372 data-points; we have 1344 training points, out of which 745 correspond to class 0 and 599 to class 1 and 28 test points. To find the most accurate representation of quantum state $\ket{\psi}_1$ (class 0), $\ket{\psi}_2$ (class 1) out of these 1344 points is the same as the difficulties discussed in Sec. \ref{Sec4B1}. Thus we can adopt the strategies discussed to resolve it; the encoding procedure from Sec. \ref{Sec4B2} where we proposed an encoding procedure to encode the 4 features of banknote dataset in 2 qubit system and from Sec. \ref{Sec4B3A} and \ref{Sec4B3K} where we propose 2 classical pre-processing methods. For proposed method on the banknote dataset we also adapt both the classical pre-processing methods, averages and k-means. From Sec. \ref{Sec4B3A} of averages, we recall Fig. \ref{QSVM_Fig7g} and Fig. \ref{QSVM_Fig7h} correspond to train 1 and train 2 circuits. Which in the proposed case implies the unitaries $U_1$, $U_2$ corresponding to quantum states $\ket{\psi}_1$, $\ket{\psi}_2$. The $U_0$ unitary corresponding to test points is obtained in similar fashion, by normalising each test point and then use the encoding procedure (Sec. \ref{Sec4B2}). Thus $A_{10}$ circuit and $A_{20}$ circuit for averages case can be seen in Fig. \ref{QSVM_Fig10a} and Fig. \ref{QSVM_Fig10b} respectively. 
Similarly From Sec. \ref{Sec4B3K} of k-means, we get $A_{10}$ circuit and $A_{20}$ circuit for k-means case as shown in Fig. \ref{QSVM_Fig10c} and Fig. \ref{QSVM_Fig10d} respectively. The $A_{10}$ and $A_{20}$ circuits shown in Fig. \ref{QSVM_Fig10a} to Fig. \ref{QSVM_Fig10d}, could easily to extended to multiple layers by repeatedly applying the unitaries $U_1$, $U_2$ and $U_0$. For our banknote case we run the $A_{10}$ and $A_{20}$ circuits for layer 1 to layers 60 on IBM quantum simulator. Line graph comparing the simulator results layers wise for both averages and k-means can be seen in Fig. \ref{QSVM_Fig10e}. We do the same for layer 1 to layers 10 on IBM real chip. Line graph comparing the real chip results layers wise for both averages and k-means can be seen in Fig. \ref{QSVM_Fig10f}.    

\textbf{Summary of results for banknote case:}
\begin{enumerate}
    \item \textbf{IBM qasm simulator:} 
    \begin{itemize}
        \item Averages; we observe a peak accuracy of \textbf{67.86\%} for layer 20 and 60.
        \item K-mean; we observe a peak accuracy of \textbf{78.6\%} for layer 4.
    \end{itemize}
    \item \textbf{IBM belem:}  
        \begin{itemize}
        \item Averages; we observe a peak accuracy of \textbf{64.3\%} for layer 2.
        \item K-mean; we observe a peak accuracy of \textbf{82.1\%} for layer 4.
    \end{itemize}
\end{enumerate}



Specification of quantum system used in this section.
\begin{enumerate}
    \item IBM qasm simulator: 32 qubits. Simulator type: General, context-aware.
    \item IBM armonk: 1 qubit; 1 quantum volume; Avg. T1: 188.86 $\mu$s; Avg. T2: 232.57 $\mu$s.
    \item IBM belem specifications: 5 qubits; 16 quantum volume; 2.5K CLOPS; Avg. T1: 93.81 $\mu$s; Avg. T2: 102.93 $\mu$s.
\end{enumerate}

\section{Conclusion \label{SecV}}

To conclude, we have investigated the quantum support vector machine (QSVM) algorithm \cite{RebentrostPRL2014,LiPRL2015} and found its technical difficulties while implementing it on current NISQ era quantum simulator and real chip. This is evident from the results for 6/9 dataset where in authors \cite{LiPRL2015} claim 100\% accuracy on NMR device whereas we got a mere accuracy of 62.5\% on IBM quantum experience for both simulator and real chip. Similarly when applied to banknote dataset also we don't seem to get expected results. This could be because of formation of an erroneous solution state, due no proper post selection after HHL algorithm. In future works, we may think of ways to do proper post-selection upon which we may get better accuracy's for the binary classification problem. In work given by Aaronson \cite{AaronsonNatP2015}, discusses more such caveats related to HHL algorithm. We also observed that there was not much change in the results with respect to $\gamma$. We would also like to mention that the differences between the QSVM circuit we proposed in Fig. \ref{QSVM_Fig3} and given \cite{LiPRL2015}. We haven't used any swap operation for FT terms, and before measurement following the swap test we have applied a H gate. While implementing QSVM algorithm for banknote dataset we proposed an encoding procedure, which can encode a general 4 feature vectors in a 2 qubit system (Fig. \ref{QSVM_Fig5}). Later part of the paper, we propose a new method which we observe performs with higher efficiency. In 6/9 case with get an 100\% accuracy for both simulator and real chip, as compare to just 62.5\% while we implement using the QSVM circuit in Fig. \ref{QSVM_Fig3}. Similarly we observe for banknote dataset a 82.1\% accuracy on real chip, and 78.6\% on simulator compare to nearly 64\% for the QSVM circuit in Fig. \ref{QSVM_Fig69}.

\section*{Acknowledgements\label{qnn_acknowledgements}}
The authors would like to thank Dr. Radhika Vathsan for her discussion and comments during the course of this paper. The authors would also like to acknowledge the support of IBMQ experience platform for providing free access to the real quantum devices and simulator. The discussions and opinions developed in this paper are only those of the authors and do not reflect the opinions of IBM or IBM QE team. The authors are also thankful to Bikash's Quantum (OPC) Pvt. Ltd. for providing the assistance throughout the project.

\end{document}